\documentclass[twocolumn, amsmath]{aastex702}
\usepackage{CJKutf8}
\usepackage{siunitx}

\begin{document}
\begin{CJK*}{UTF8}{min}

\title{Dynamical formation of Gaia neutron star binaries in star clusters as a phase-dependent process}

\author[gname=Ryutaro,sname=Iwakura, orcid=0009-0009-8049-2968]{Ryutaro Iwakura (\CJKfamily{ipxg}岩倉 龍太郎)}
\affiliation{Department of Planetology, Graduate School of Science, Kobe University, 1-1 Rokkodai-cho, Nada-ku, Kobe, Hyogo 657-8501, Japan}
\affiliation{School of Physics and Astronomy, Sun Yat-sen University, Daxue Road, Zhuhai 519082, China}
\email[show]{riwakura@stu.kobe-u.ac.jp}

\author[gname=Long,sname=Wang, orcid=0000-0001-8713-0366]{Long Wang (\CJKfamily{gbsn}王龙)}
\affiliation{School of Physics and Astronomy, Sun Yat-sen University, Daxue Road, Zhuhai 519082, China}
\affiliation{CSST Science Center for the Guangdong-Hong Kong-Macau Greater Bay Area, Zhuhai 519082, China}
\email{wanglong8@sysu.edu.cn}

\author[gname=Ataru,sname=Tanikawa, orcid=0000-0002-8461-5517]{Ataru Tanikawa (\CJKfamily{ipxg}谷川 衝)}
\affiliation{Center for Information Science, Fukui Prefectural University, 1-1 Matsuoka Kenjojima, Eiheiji-cho, Fukui 910-1195, Japan}
\email{tanik@g.fpu.ac.jp}

\author[gname=Takayuki R.,sname=Saitoh, orcid=0000-0001-8226-4592]{Takayuki R. Saitoh (\CJKfamily{ipxg}斎藤 貴之)}
\affiliation{Department of Planetology, Graduate School of Science, Kobe University, 1-1 Rokkodai-cho, Nada-ku, Kobe, Hyogo 657-8501, Japan}
\email{saitoh@people.kobe-u.ac.jp}

\author[gname=Yoko,sname=Funato, orcid=0000-0002-6992-7600]{Yoko Funato (\CJKfamily{ipxg}船渡 陽子)}
\affiliation{General Systems Studies, Graduate School of Arts and Sciences, The University of Tokyo, Komaba, Meguro, Tokyo 153-8902, Japan}
\email{funato@system.c.u-tokyo.ac.jp}

\author[gname=Junichiro,sname=Makino, orcid=0000-0002-0411-4297]{Junichiro Makino (\CJKfamily{ipxg}牧野 淳一郎)}
\affiliation{Department of Planetology, Graduate School of Science, Kobe University, 1-1 Rokkodai-cho, Nada-ku, Kobe, Hyogo 657-8501, Japan}
\affiliation{Preferred Networks, Inc., 1-6-1 Otemachi, Chiyoda-ku, Tokyo 100-0004, Japan}
\email{jmakino@people.kobe-u.ac.jp}

\correspondingauthor{Ryutaro Iwakura}

\begin{abstract}
Gaia has discovered dormant compact-object binaries in which the compact object shows no observable interaction with its luminous companion.
These systems, hereafter Gaia BHs and Gaia NSs for black hole (BH) and neutron star (NS) companions, provide a new probe of compact-object binary formation.
These detached systems are difficult to produce through standard isolated binary evolution alone, and their formation pathways remain unclear.
Previous $N$-body studies have shown that Gaia BH-like objects are formed efficiently in star clusters, while Gaia NS-like objects are not.
This has been attributed to the BH subsystem dominating the cluster core.
We performed direct $N$-body simulations of $10^4\ \mathrm{M}_\odot$ star clusters in a Galactic tidal field and investigated how NS--main-sequence (NS--MS) binaries form during the late BH-depleted phase.
We found that NS--MS binaries satisfying Gaia-sensitive criteria form only after BH depletion. 
NSs are present in the core throughout the evolution, but the strong encounters there are monopolized by the BH subsystem until it is depleted.
Our results show that the dynamical formation of Gaia-sensitive NS--MS systems is not prevented in star clusters, but delayed relative to Gaia BH-like systems.
\end{abstract}


\section{Introduction}
The Gaia mission has enabled the detection of non-accreting compact-object binaries through high-precision astrometry \citep{GaiaDR3, El-Badry2024}.
To date, three non-accreting black holes \citep[BHs;][]{Chakrabarti+2023, El-Badry+2023BH1, El-Badry+2023BH2, Tanikawa+2023GaiaBH, GaiaBH3} and 21 candidate neutron stars \citep[NSs;][]{El-badry+2024NS1, El-Badry+2024NS21} have been reported in binaries with main-sequence (MS) and giant companions.

Hereafter, we refer to these astrometric BH and NS binaries identified through Gaia as ``Gaia BH'' and ``Gaia NS'', respectively, and collectively as ``Gaia compact binaries.''
These systems are characterized by relatively long orbital periods, and therefore do not currently exhibit significant mass transfer from their stellar companions.
The formation pathways of Gaia compact binaries remain poorly understood, especially because their wide, detached orbital configurations are difficult to reproduce through standard isolated binary evolution \citep{Kotko+2024, Gilkil&Mazeh2024, Kruckow+2024, Mapelli+2026}.

In the usual evolution of isolated binaries, as the compact-object progenitor expands, the system undergoes Roche-lobe overflow and often common-envelope (CE) evolution, which generally leads to substantial orbital shrinkage or merger \citep[e.g.,][]{Ivanova+2013, Tauris+2017}.
For NS binaries, this difficulty is compounded by supernova mass loss and strong natal kicks, which can readily disrupt the binary even if it survives the earlier interaction phases \citep{Hobbs+2005}.
Recently, \citet{Hirai+2024} proposed an isolated-binary scenario that combines NS natal kicks with the NS rocket mechanism and can form wide, detached NS binaries. 
This scenario can reproduce the observed orbital properties \citep{Baibhav+2026} and Galactic trajectories \citep{Disberg+2026} of Gaia NSs, although the physical conditions required for an effective rocket remain uncertain \citep{Baibhav+2026}.

Dynamical interactions in star clusters have also been studied as an alternative formation channel for Gaia compact binaries.
In particular, for Gaia BHs, both observational and theoretical studies support their cluster origin: Gaia BH3 is likely to be associated with the ED-2 stream \citep{Balbinot+2024}, and recent $N$-body and Monte Carlo studies have shown that Gaia BH-like binaries can be produced through exchange interactions and dynamical capture in star clusters \citep{Rastello+2023, DiCarlo+2024, MarinPina+2024, Tanikawa+2024, Fantoccoli+2025, MarinPina+2025}. 
Furthermore, the detached BH binary candidates in globular clusters discovered prior to the Gaia BHs \citep{Giesers+2018, Giesers+2019} can also be explained by dynamical interactions \citep{Kremer+2018}.

In contrast, the dynamical formation of Gaia NSs seems to be difficult.
\citet{Tanikawa+2024NS} studied open clusters with initial
masses of $\sim 10^3\ \mathrm{M}_\odot$ and found that Gaia NSs form far less efficiently than Gaia BHs.
Even when core-collapse supernova (CCSN) natal kicks were switched off, dynamically formed Gaia NSs remained rare.
They attributed this difficulty to the fact that NS masses are comparable to or lower than those of the surrounding MS stars.
They also noted that the main-sequence lifetime of a $1.4\ \mathrm{M}_\odot$ star is a few Gyr, longer than the lifetime of the open clusters they considered.
\citet{Fantoccoli+2025} found no dynamically formed NS--star binaries in their simulations.
They argued that this is because BHs dominate the core of clusters until the end of the simulations, and they make companion acquisition by NSs difficult.

However, this early-phase picture might not be complete.
Recent long-term $N$-body simulations by \citet{Banerjee2026} have shown that young massive and open cluster models simulated until BH depletion can yield Gaia NSs with parameters consistent with Gaia candidates.
Before BH depletion, when the subsystem made of BHs occupies the cluster core, only BHs can undergo strong dynamical interactions, suppressing the dynamical formation and evolution of low-mass binaries \citep{Wang2022}. 
Therefore the dynamical formation of Gaia NSs depends sensitively on the evolutionary phase of the host cluster.
The implications of this late-time cluster evolution for Gaia NS formation require further investigation.
In particular, it is still unclear how the transition from the BH-dominated phase to the BH-depleted phase affects the ability of NSs to participate in strong encounters and exchange interactions in the core.

The low formation efficiency of Gaia NSs reported in previous studies can be understood from two related effects.
First, NSs generally receive natal kicks at birth, so their retention fraction in star clusters is low unless the contribution of low-kick channels is significant \citep{Hobbs+2005, Verbunt+2017}.
More importantly for the present study, even retained NSs have few opportunities to acquire companions while the BH subsystem is present.
This suppression can be understood in terms of Spitzer instability, in which the most massive objects form a dynamically decoupled subsystem \citep{Spitzer69, BreenHeggie2013, Banerjee2018, Fragione+2018, Ye+2020, Kremer+2020,Wang2020}.
The BH subsystem therefore dominates the strong encounters in the core, and the NSs that do reach the core are unlikely to be the ones that undergo them.

In this paper, we investigate the hypothesis that the dynamical channel for Gaia NS-like binaries is not closed by the position of NSs in the cluster, but by the presence of the BH subsystem, and that it opens once the BHs are depleted.
As BHs are progressively ejected and the MS turn-off mass decreases, NSs become the most massive components in the cluster, so that they both participate in and survive as the retained member of exchange encounters in the core.
Testing this hypothesis requires direct $N$-body simulations that follow the long-term dynamical evolution of star clusters beyond the BH-dominated phase.
We therefore performed direct $N$-body simulations of $10^4\ \mathrm{M}_\odot$ star clusters in a Galactic tidal field up to $7$ Gyr.

The structure of this paper is as follows.
In Section 2, we describe the numerical method and the initial conditions.
In Section 3, we present our results.
In Section 4, we discuss the implications and limitations of our results, and compare them with previous studies.
Finally, in Section 5, we summarize our work and describe possible future work.

\section{Methods}
\subsection{Numerical code}
We use the direct $N$-body code \textsc{petar} \citep{petar}.
\textsc{petar} adopts the Particle-Particle Particle-Tree scheme \citep{Oshino+2011}, which divides the gravitational force into a long-range component and a short-range component.
The long-range force is calculated with the Barnes-Hut tree algorithm \citep{BarnesHut86} and is integrated with a second-order leapfrog scheme.
The short-range force is computed by the direct method and is integrated with the fourth-order Hermite scheme \citep {MakinoAarseth92}.
Few-body dynamics, such as binary systems and hyperbolic encounters, are handled by Slow-down Algorithmic Regularization \citep[\textsc{sdar};][]{sdar}.
This design integrates the close encounters and few-body dynamics accurately and keeps the overall computational cost low.
\textsc{petar} is parallelized with Framework for Developing Particle Simulations \citep[\textsc{fdps};][]{Iwasawa+2016,Iwasawa+2020} and supports GPU acceleration.
For single and binary stellar evolution, \textsc{petar} uses the \textsc{bse} package \citep{Hurley+2000, Hurley+2002, Banerjee+2020}.
We model the external Galactic potential with MWPotential2014 of \textsc{galpy} \citep{galpy}.

\subsection{Initial profiles of clusters}
Our objective is to investigate whether Gaia NS-like binaries can form dynamically during the late evolutionary phase of star clusters, after the depletion of BHs from the core of the cluster.
To address this problem, our cluster models must satisfy two key requirements:
(i) early depletion of stellar-mass BHs through dynamical interactions, and
(ii) long-term survival of the cluster over several Gyr.
We therefore construct initial conditions that satisfy both of these requirements.

We generate all initial conditions with \textsc{McLuster} \citep{Kuepper+2011, Wang+2019}.
The initial density profile follows the Plummer model \citep{Plummer11}.
The initial cluster mass is $10^4\ \mathrm{M}_\odot$.
This mass lies between the upper end of the mass range of massive open clusters and the lower end of that of young massive clusters (YMCs).

To ensure the efficient early dynamical evolution and depletion of BHs, we adopt a compact initial configuration with a half-mass radius of $r_\mathrm{h} = 0.5$ pc.
More extended clusters evolve more slowly and the depletion time of BHs is longer.
They also dissolve earlier, and we cannot follow the late evolutionary phase of the cluster.

All clusters are placed on circular orbits at the Galactocentric distance of $12$ kpc, so that the lifetime of the cluster is longer than the depletion time of BHs.
This is essential for testing our hypothesis.
At smaller Galactocentric radii, strong tidal stripping leads to rapid cluster dissolution.
We consider two metallicities, $Z=0.02$ and $Z=0.002$, to examine the sensitivity of our results to the models for stellar evolution and compact object formation.

The initial mass function follows that of \citet{Kroupa2001}, with the lower and upper mass limits of $0.08\ \mathrm{M}_\odot$ and $150\ \mathrm{M}_\odot$, respectively.
We include primordial binaries in the following way.
For massive stars ($m \geq 5\ \mathrm{M}_\odot$), we adopt the binary properties reported by \citet{Sana+2012} and \citet{Oh+2015}. 
The binary fraction is unity.
For lower-mass stars ($m < 5\ \mathrm{M}_\odot$), binary parameters are drawn from the distributions of \citet{Kroupa95a,Kroupa95b} and \citet{Belloni+2017}, and the overall binary fraction is $0.2$.
For each of the two metallicities, we generated 10 independent realizations with different seeds. 
Our fiducial model set therefore consists of $20$ cluster simulations.

\subsection{Compact object formation and natal kicks}
In this section, we describe the models of compact-object formation implemented in \textsc{bse}.

The CE phase largely determines the orbital properties of compact binaries.
In \textsc{bse}, CE evolution is treated with the standard energy formalism and the efficiency parameter $\alpha_\mathrm{CE}$ \citep{Webbink84}.
In this study, we adopt $\alpha_\mathrm{CE} = 1$ and the envelope binding-energy parameter $\lambda_\mathrm{CE}$ is fixed to  $0.5$.
These choices prevent wide primordial binaries from surviving CE evolution.
As a result, isolated binary evolution does not produce detached compact binaries in our models.
We can therefore isolate the contribution of the dynamical formation channel, which is the primary target of the present study.

Natal kicks strongly affect the retention of compact objects in star clusters, especially that of NSs.
In the present study, we consider three natal kick prescriptions for NS formation.
For NSs formed via CCSNe, we adopt a Maxwellian velocity distribution with a dispersion of  $\sigma_\mathrm{ccsn} = 265$ \unit{km.s^{-1}} \citep{Hobbs+2005}.
In this channel, the kick velocity is reduced according to the fallback prescription of \citet{Fryer+2012}.
The same prescription is also applied to BH natal kicks.
Because the typical CCSN kick is much larger than the escape velocity of our clusters, most NSs formed through this channel escape from the cluster.

The other two channels produce NSs with low natal kicks.
For electron-capture supernovae (ECSNe), we adopt a Maxwellian velocity distribution with $\sigma_\mathrm{ecsn} = 3$ \unit{km.s^{-1}}, which represents the weakly anisotropic explosion model \citep{Podsiadlowski+2004,GessnerJanka2018}.
We also consider the accretion-induced collapse (AIC) channel, for which we assume zero natal kick \citep{Baron+87, NomotoKondo91, WoosleyBaron92, Fryer+99, Dessart+2006, Abdikamalov+2010, Wang2018, WangLiu2020, LongoMicchi+2023, Mori+2025}.
We consider two types of AIC events: (i) an oxygen-neon white dwarf
(ONeWD) accretes material from its companion and exceeds the Chandrasekhar limit,
and (ii) two WDs merge through gravitational-wave-driven inspiral.
The former channel is already implemented in \textsc{bse}.
For the latter channel, we adopt the WD-merger AIC prescription of \citet{Tanikawa+2024NS}.

We note that \citet{Disberg&Mandel2025} recently showed that the distribution of \citet{Hobbs+2005} overestimates the median NS natal kick velocity by $\sim 50$\%. 
Adopting the revised distribution would therefore increase the retention of CCSN-formed NSs. 
Nevertheless, typical CCSN kicks would remain an order of magnitude larger than the characteristic escape velocity of our clusters ($\approx 15$ \unit{km.s^{-1}}). 
We therefore expect the retained NS population to remain dominated by the low-kick ECSN and AIC channels, and our qualitative conclusions to remain unchanged.

\section{Results}
In this section, we examine how depletion of BHs changes the dynamical environment of retained NSs and enables the late formation of Gaia NS-like binaries.
We first present the long-term evolution of the cluster structure and of the compact-object populations.
Dynamically formed BH--MS and NS--MS binaries appear at different epochs.
We then characterize the orbital properties of the resulting compact-object binaries, and trace the dynamical formation pathway of a representative Gaia NS-like system.

\subsection{The dynamical evolution of clusters}
\label{sec:dynamical_structure}

\begin{figure*}
    \includegraphics[width=\textwidth]{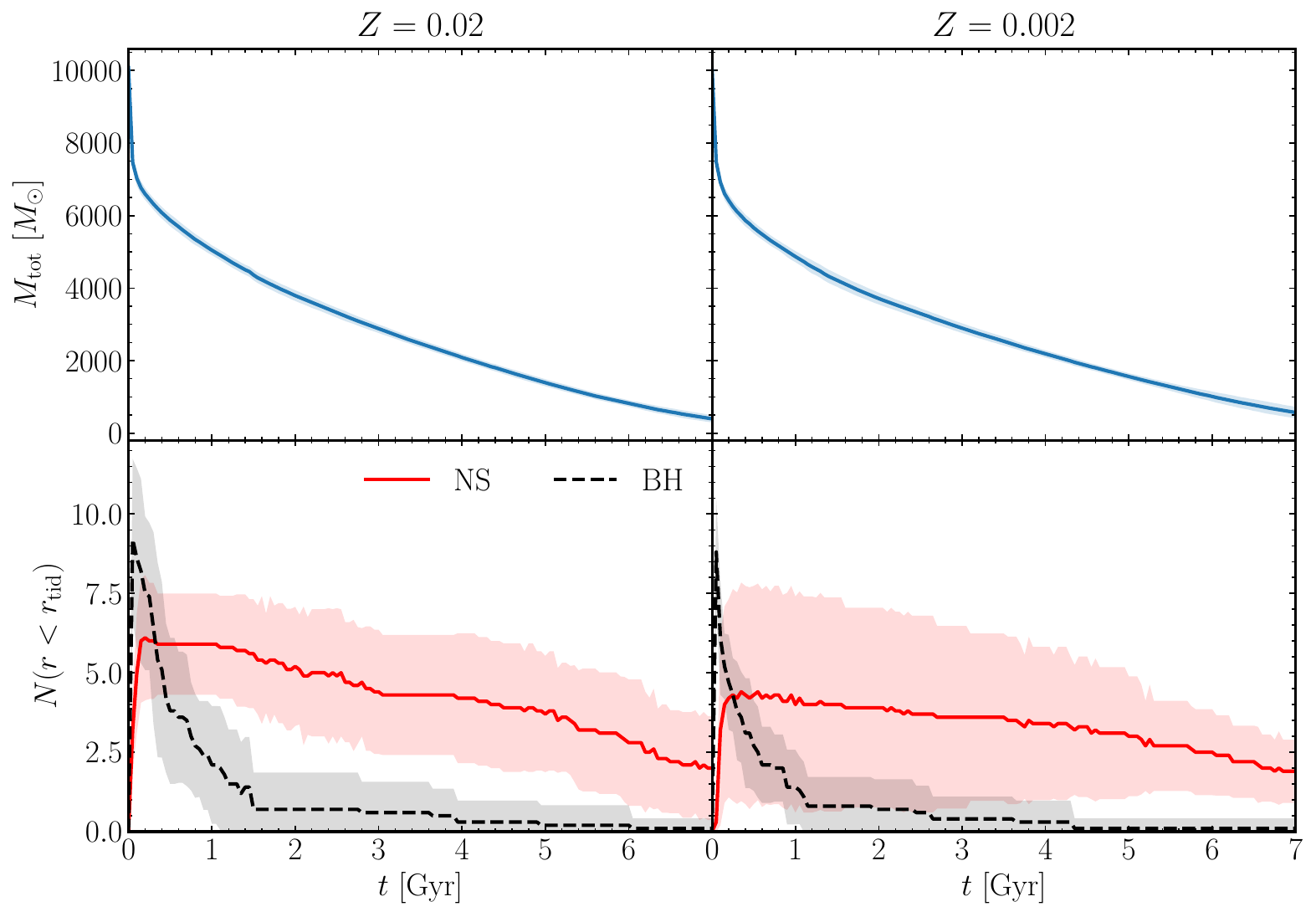}
    \caption{Evolution of the cluster mass and compact-object populations for the $Z=0.02$ (left) and $Z=0.002$ (right) models.
    The upper panels show the stellar mass enclosed within the instantaneous tidal radius.
    The lower panels show the numbers of BHs and NSs within the instantaneous tidal radius.
    Solid lines indicate the mean values over the ten realizations, while the shaded regions represent the standard deviations.
    The origin is the density center defined by \citet{Casertano&Hut85}, and the same definition is adopted throughout this paper.}
    \label{fig:num_of_compact_objects}
\end{figure*}

Figure \ref{fig:num_of_compact_objects} shows the long-term evolution of the cluster mass and the retained compact-object populations for the two metallicities. 
For both cases, the cluster mass decreases steadily because of stellar evolution and tidal mass loss. 
In contrast, the BH population declines rapidly during the first $\sim1$--$2$ Gyr and is almost depleted thereafter. 
In this paper, we define a cluster as BH-depleted when fewer than two BHs remain within its tidal radius.
The number of NSs within the tidal radius shows large variations, but on average several NSs remain for several Gyr.
The evolution of each individual realization is shown in Appendix \ref{appendix}.

We can see that a fair number of NSs are retained in the cluster and only slowly escape from the cluster.
These retained NSs participate in the dynamical formation and evolution of binaries once the BHs have been depleted, resulting in the formation of Gaia NS-like systems.

\begin{figure*}
    \includegraphics[width=\textwidth]{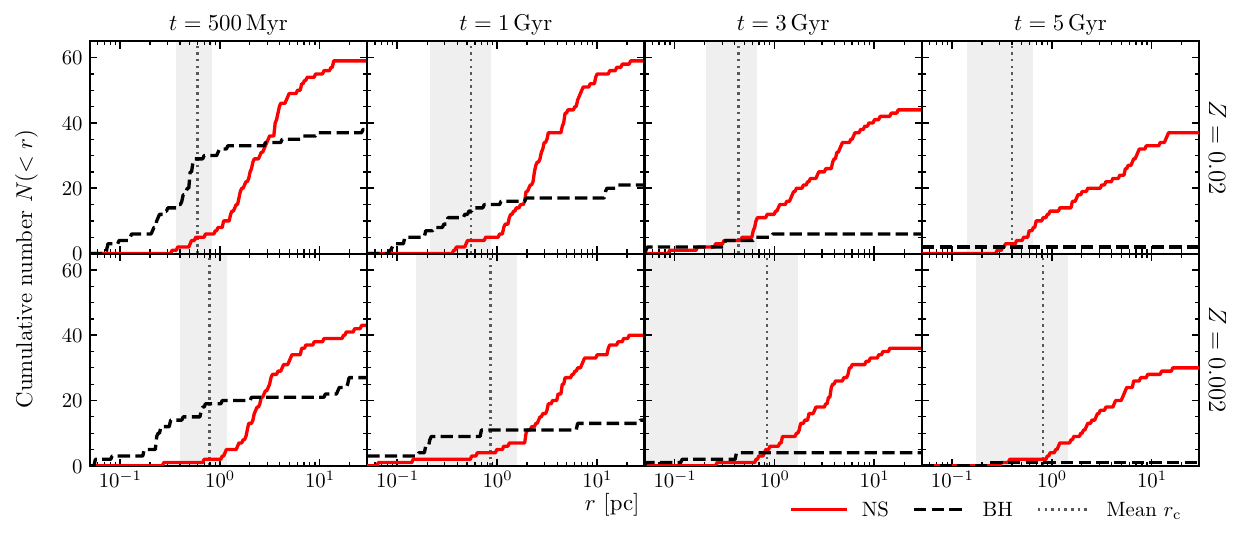}
    \caption{Cumulative radial distributions of BHs and NSs for the two metallicities.
    The distributions are obtained by combining all objects from the ten realizations at each metallicity.
    The dotted vertical lines indicate the mean core radii over the ten realizations.
    The gray shaded regions show the corresponding ranges of one standard deviation about the means.}
    \label{fig:radial_distribution}
\end{figure*}

Figure \ref{fig:radial_distribution} shows the cumulative radial distribution of BHs and NSs for each metallicity at four epochs.
At $t=500$ Myr, most of the retained BHs are inside the core. 
The distributions at $t=1$ Gyr are similar.
NSs are not completely absent from the core, but only a few are found within $r_\mathrm{c}$, an order of magnitude fewer than BHs, even though the total number of retained NSs is larger.

After almost all BHs have been depleted ($t=3$ and $5$ Gyr), the inner part of the NS distribution is not significantly different from that at the earlier epochs, and shows no discontinuous change across BH depletion.
The NSs present in the core can then take part in the dynamical formation and evolution of binaries.
A large fraction of the NSs, however, still lie outside the core at these epochs.
We note that $r_\mathrm{c}$ differs from realization to realization and increases with time, so the dotted lines are representative values rather than a threshold that applies to every run.

\begin{figure*}
    \includegraphics[width=\linewidth]{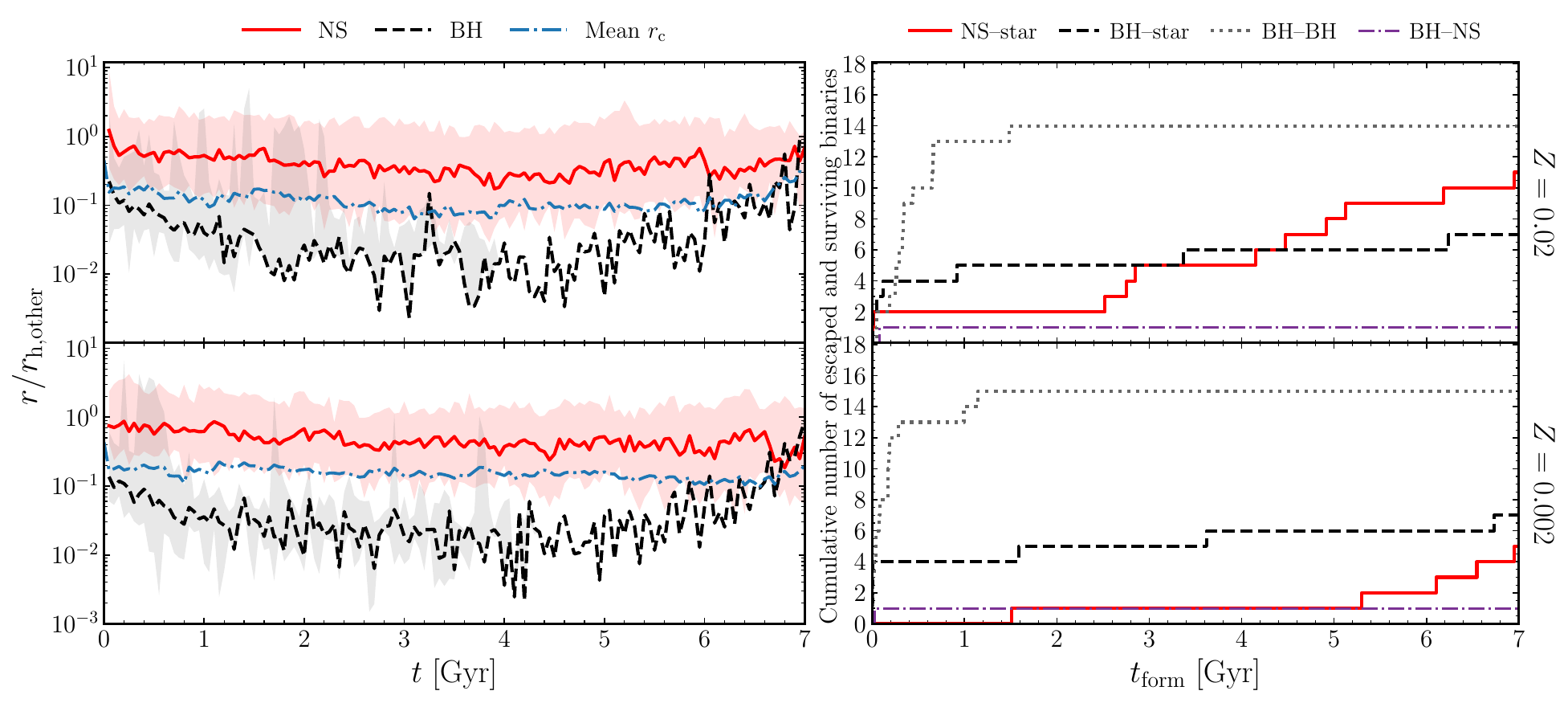}
    \caption{Left panels: Time evolution of the Lagrangian radii of NSs and BHs, computed from the NSs and BHs combined over the ten realizations at each metallicity.
    The thick curves show the 50\% Lagrangian radii, while the shaded regions represent the range from the 10\% to the 90\% Lagrangian radii.
    The dash-dotted line shows the mean core radius over the ten realizations.
    Right panels: Cumulative numbers of dynamically formed compact-object binaries that escape from the cluster or are present as bound pairs at the final snapshot at $t=7$ Gyr, shown as functions of their formation time $t_\mathrm{form}$.
    The counts are summed over the ten realizations at each metallicity and separated by binary type.
    The sample selection and the definition of $t_\mathrm{form}$ are described in Section \ref{sec:dynamical_structure}.
    The upper and lower rows show the $Z=0.02$ and $Z=0.002$ models, respectively.}
    \label{fig:lagrangian_radii}
\end{figure*}

The left panels of Figure \ref{fig:lagrangian_radii} show the time evolution of the 10\%, 50\%, and 90\% Lagrangian radii of the BH and NS populations, together with the mean core radius.
Because each realization contains only a few NSs, we normalize the distance of each NS and BH by the half-mass radius of all other cluster members in its own realization, and compute the Lagrangian radii from the NSs and BHs combined over the ten realizations at each metallicity.
Throughout most of the evolution, the 50\% Lagrangian radius of BHs remains smaller than the core radius. 
As the most massive objects in the cluster, BHs preferentially sink toward the core through mass segregation.

During the first $\sim 1$ Gyr, the 10\% Lagrangian radius of the NSs is comparable to or larger than the core radius, and at later times it lies below the core radius more continuously.
This gradual shift does not by itself imply that the NSs sink into the core after BH depletion.
The Lagrangian radii are computed from the NSs retained within the instantaneous tidal radius, and this number decreases from several to about two over the run (Figure \ref{fig:num_of_compact_objects}), so the same few NSs in the core correspond to a progressively larger fraction of the population as the outer NSs are tidally stripped.
The NS population never reaches the strong central concentration of the BHs, and its Lagrangian radii show no discontinuous change at the epoch of BH depletion.
The innermost part of the NS population occupies the cluster core both before and after the BH subsystem disappears, and these NSs can participate in dynamical interactions with binaries in the core.

The right panels of Figure \ref{fig:lagrangian_radii} show when the selected dynamically formed compact-object binaries are assembled.
We include binaries that are identified as escapers or are present as bound pairs at the final snapshot at $t=7$ Gyr.
We define the formation time $t_\mathrm{form}$ as the start of the last continuous interval in which the two components remain bound to each other up to escape or the final snapshot.
Here ``star'' denotes any companion that is neither a BH nor an NS, and therefore includes WDs as well as MS and giant stars.
No cut on orbital period is applied.
The NS--star and BH--star counts are larger than those of the corresponding MS-only samples in Section \ref{sec:prop_of_binaries} because companions are not restricted to MS stars.

BH--BH binaries in this sample form during the first $\sim 2$ Gyr, whereas most NS--star binaries form later.
The timing of NS--star formation differs between the two metallicities.
For $Z=0.02$, the cumulative NS--star count reaches two at early times and begins to increase again at $\sim 2.5$ Gyr.
For $Z=0.002$, the count remains at one until $\sim 5$ Gyr, with most systems forming thereafter.
The selected sample contains one BH--NS binary at each metallicity and no NS--NS binaries.
Thus, NS--star binaries in this sample are assembled predominantly after the typical epoch of BH depletion.
Together with the radial distributions, this trend supports our interpretation that the BH subsystem suppresses dynamical NS-binary formation even when some NSs are already present in the core.

\subsection{Properties of dynamically formed compact binaries}
\label{sec:prop_of_binaries}
We first define the binary classes used in this section.
We define a BH--MS or NS--MS binary as dynamically formed if the compact object acquired its present MS companion through at least one dynamical interaction inside the cluster.
This definition does not require the compact object itself to have formed as a single star; it may originate from a primordial binary.
The essential requirement is that the present compact-object--MS pair was not inherited from the primordial binary population.

For comparison with Gaia astrometric compact-object binaries, we define Gaia-sensitive systems as dynamically formed BH--MS or NS--MS binaries that satisfy two criteria: (i) they have orbital periods in the range $10^2 \leq P/{\rm day} \leq 10^4$, and (ii) they are no longer associated with a bound star cluster, either because they are explicitly identified as escapers or because they are recorded at the final snapshot after the host cluster has effectively dissolved.
We do not impose additional cuts on companion mass or eccentricity, because our purpose is not to reproduce the Gaia selection function, but to identify the dynamical formation pathway of dormant compact-object--MS binaries.

We record BH/NS--MS binaries either at the time they are explicitly identified as escapers or, if they remain in the simulation until the end, at the final snapshot.
For escaped binaries, the orbital elements are measured at the time of escape.
For final-snapshot binaries, the orbital elements are measured at 7 Gyr.
By this time, our cluster models have lost more than 95\% of their initial mass and are therefore effectively dissolved.
The escape criterion based on the tidal radius becomes ill-defined at this stage, and we do not classify final-snapshot binaries as escapers.
Instead, we treat them as binaries associated with the dissolved cluster remnant.
The formation time $t_\mathrm{form}$ is defined as in Section \ref{sec:dynamical_structure}, here applied to the BH/NS--MS pairs.
The interval ends either at escape or at the final snapshot.

We find 20 dynamically formed compact-object--MS binaries in the 20 realizations.
Eleven of them are BH--MS and nine are NS--MS binaries.
The Gaia-sensitive criteria select four BH--MS systems and three NS--MS systems.
All Gaia-sensitive binaries are explicitly identified as escapers before the final snapshot.
Our sample is too small to give a robust formation efficiency.
It nevertheless shows that dynamically formed NS--MS systems reach the Gaia-sensitive range in our simulations.

\begin{figure*}
    \includegraphics[width=\linewidth]{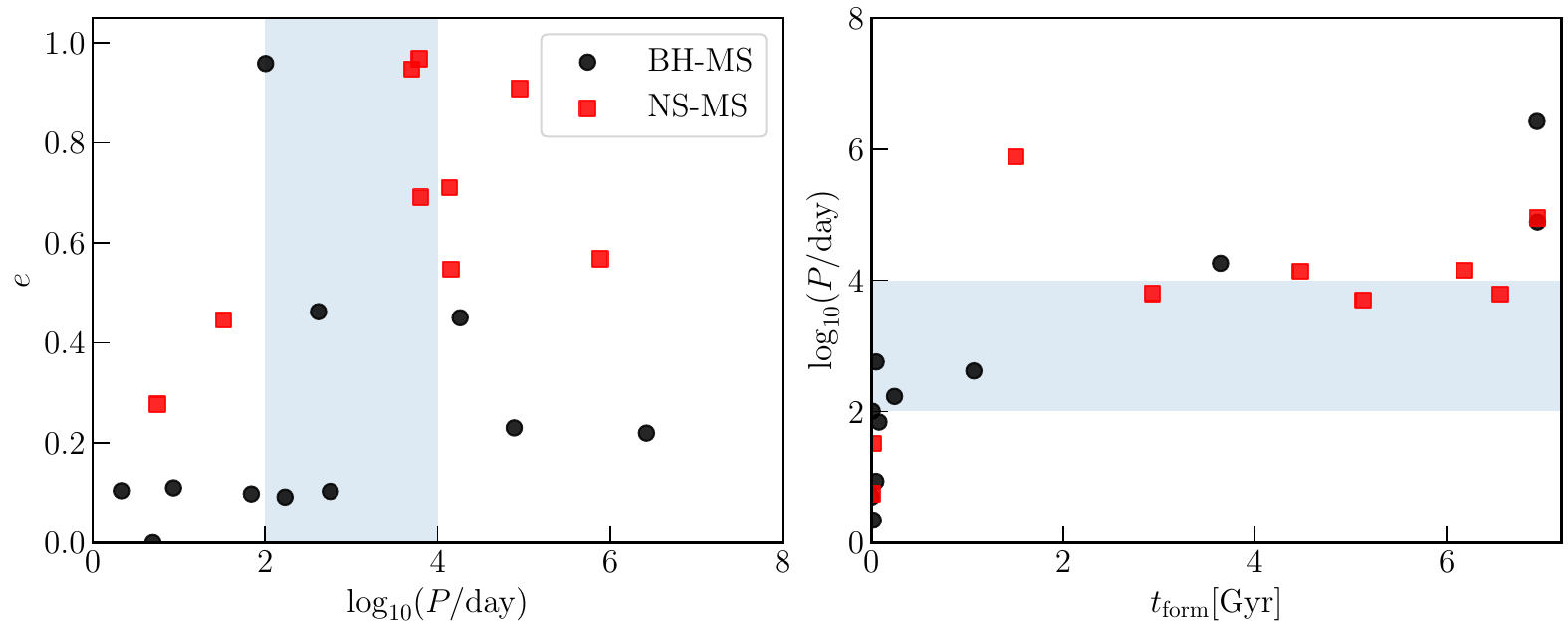}
    \caption{
    Orbital properties and formation times of dynamically formed compact object--MS binaries.
    The shaded region indicates the Gaia-sensitive period range, $10^2 \leq P/\mathrm{day} \leq 10^4$.
    The formation time $t_\mathrm{form}$ is defined as in Section \ref{sec:dynamical_structure}.
    For escaped binaries, the orbital elements are measured at the time of escape; for final-snapshot binaries, they are measured at 7 Gyr.
    }
    \label{fig:compact_pe_scatter}
\end{figure*}

Figure \ref{fig:compact_pe_scatter} shows the orbital properties and formation times of the dynamically formed compact-object--MS binaries.
No period cut is applied to the plotted sample.
The shaded region indicates the Gaia-sensitive period range, $10^2 \leq P/\mathrm{day} \leq 10^4$.
Two of the nine NS--MS binaries form and escape within the first $\sim 30$ Myr.
Their orbital periods are below $10^2$ days, and one of them has a massive MS companion of $\sim 8\ \mathrm{M}_\odot$.
These early systems appear in the plotted sample, but they do not satisfy our Gaia-sensitive criteria.
They are distinct from the late-time NS--MS binaries discussed below.

The left panel of Figure \ref{fig:compact_pe_scatter} shows that dynamically formed compact-object--MS binaries span a wide range of orbital periods and eccentricities.
The NS--MS binaries show a tendency toward longer periods and higher eccentricities than most BH--MS binaries, although the sample size is small.
In particular, the Gaia-sensitive NS--MS binaries lie near the long-period end of the adopted Gaia-sensitive range and have high eccentricities.
Late-time dynamical NS--MS formation in these models therefore produces wide and eccentric systems.
It does not populate the short-period or low-eccentricity regime efficiently.

The formation-time distribution provides a direct connection to the global cluster evolution discussed in Section \ref{sec:dynamical_structure}.
BH--MS binaries are preferentially assembled during the early phase, when the cluster core is still dominated by retained BHs.
By contrast, the Gaia-sensitive NS--MS binaries can be assembled after the depletion of the BH subsystem, when the strong encounters in the core are no longer monopolized by BHs and NSs can participate in exchange interactions.
The resulting NS--MS binaries are often wide, because they form in the late evolutionary phase when the cluster has already expanded and the encounter rate is reduced.
Thus, late-time cluster evolution opens the dynamical channel for NS--MS formation, but subsequent hardening to $P \lesssim 10^3$ days may be inefficient in these $10^4\ \mathrm{M}_\odot$ cluster models.

\begin{table*}
    \caption{
    Properties of Gaia-sensitive dynamically formed NS--MS binaries in our simulations.
    The orbital elements are measured at the selected escaped state.
    Here, $r_{\rm NS}$ and $r_{\rm c}$ are the distance of the NS from the cluster density center and the cluster core radius, respectively, at the last available snapshot before $t_{\rm form}$.
    }
    \label{tab:ns_info}
    \centering
    \begin{tabular}{cccccccc}
    \hline
    $P\,[\mathrm{day}]$
    & $e$
    & $m_{\rm NS}\,[M_\odot]$
    & $m_{\rm MS}\,[M_\odot]$
    & $t_{\rm form}\,[\mathrm{Gyr}]$
    & $r_{\rm NS}\,[\mathrm{pc}]$
    & $r_{\rm NS}/r_{\rm c}$
    & $Z$ \\
    \hline
    6127 & 0.968 & 1.26 & 0.27 & 6.557 & 0.058 & 0.86 & 0.002 \\
    4980 & 0.947 & 2.35 & 2.30 & 5.128 & 0.233 & 0.71 & 0.02  \\
    6308 & 0.691 & 1.87 & 1.01 & 2.928 & 0.077 & 0.33 & 0.02  \\
    \hline
    \end{tabular}
\end{table*}

Table \ref{tab:ns_info} summarizes the orbital properties and the formation times of the three Gaia-sensitive NS--MS binaries.
All three systems form after the depletion of the BH subsystem.
This supports our interpretation that NS--MS formation occurs preferentially in the late BH-depleted phase.
At the last snapshot before binary formation, all three NSs are located within the core radius of their host clusters at that time. 
Two are the innermost NSs in their respective clusters, while the remaining one is the second innermost.

One system has a relatively massive MS companion of $m_\mathrm{MS}=2.30\ \mathrm{M}_\odot$, which may appear unusual at these late times.
This companion is a rejuvenated MS star.
It formed through a merger of two MS stars, and it is analogous to a blue straggler star.
Thus, even in the late evolutionary phase of the cluster, NSs with masses of order $2\ \mathrm{M}_\odot$ or below can acquire MS companions through dynamical interactions.
The same system also contains the most massive NS in our sample ($m_{\rm NS}=2.35\,\mathrm{M}_\odot$).
This NS formed through the merger of an ONeWD and a carbon-oxygen WD (COWD), followed by AIC.

Previous studies have reported that Gaia NS-like binaries are formed less efficiently than Gaia BH-like binaries in star clusters \citep{Tanikawa+2024NS, Fantoccoli+2025}.
Our results do not contradict this conclusion, because our simulation set is too small to determine a robust formation efficiency.
The apparent scarcity of Gaia NS-like systems is instead partly a matter of timing.
During the early BH-dominated phase, the strong encounters in the core are almost entirely taken up by BHs, even though NSs are present there.
After BH depletion, this dynamical suppression is removed, and NSs can acquire MS companions through exchange interactions.

We therefore interpret Gaia NS formation as a phase-dependent process.
In our models, Gaia-sensitive NS--MS binaries can form dynamically, but their production is delayed relative to BH--MS binaries and tends to occur in a late-time cluster environment where additional hardening is limited.
A broader survey over cluster mass, radius, and metallicity is required to determine whether more massive or denser clusters can more efficiently populate the short-period Gaia NS regime.

\subsection{Formation pathway of a Gaia NS-like binary}

\begin{figure}
    \begin{center}
    \includegraphics[width=\columnwidth]{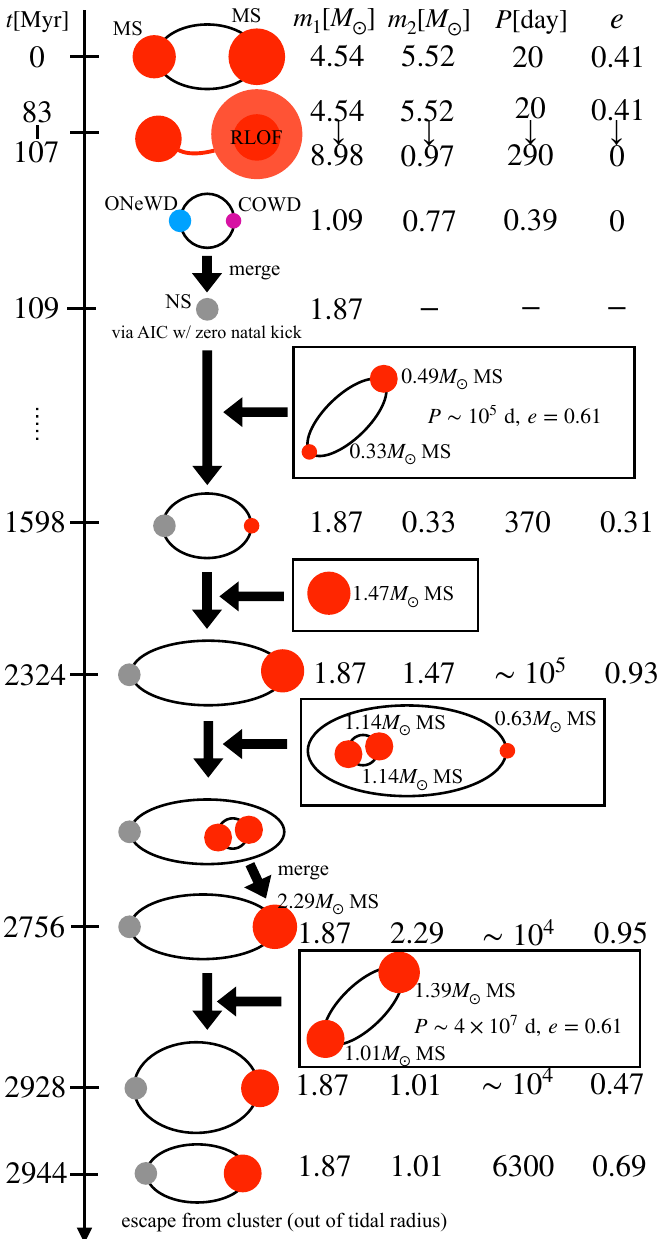}
    \end{center}
    \caption{Schematic illustration of the formation pathway of a Gaia NS-like binary found in the $Z=0.02$ model.}
    \label{fig:formation}
\end{figure}

Figure \ref{fig:formation} shows a representative formation pathway of a Gaia NS-like binary in a $Z=0.02$ model.
This example highlights three important points.
First, the NS is formed through a low-kick channel, namely the AIC channel, which allows it to be retained in the cluster.
Second, the NS remains single for about $1.5$ \unit{Gyr} after its formation and does not participate in dynamical exchanges until the BH subsystem is almost completely depleted.
Third, a single exchange does not produce the final wide NS--MS binary.
The binary is the result of a series of dynamical interactions, which include temporary pairings and chaotic few-body encounters.

The NS originates from a primordial binary composed of two MS stars with masses of $m_1=4.54\ \mathrm{M}_\odot$ and $m_2=5.52\ \mathrm{M}_\odot$.
Between $t \simeq 83$ and $107$ \unit{Myr}, stable Roche-lobe overflow increases $m_1$ to $8.98\ \mathrm{M}_\odot$ and reduces $m_2$ to $0.97\ \mathrm{M}_\odot$.
The orbit circularizes during this mass-transfer episode.
Subsequent stellar and binary evolution produces a close binary composed of an ONeWD and a COWD.
At $t \simeq 109$ \unit{Myr}, the two WDs merge and form a $1.87\ \mathrm{M}_\odot$ NS through AIC with zero natal kick.
The NS is retained in the cluster and remains single for $\sim 1.5$ \unit{Gyr}.
The BH subsystem is depleted during this interval.

At $t \simeq 1.6$ \unit{Gyr}, a close encounter with a binary composed of $0.33$ and $0.49\ \mathrm{M}_\odot$ MS stars temporarily pairs the NS with the $0.33\ \mathrm{M}_\odot$ star.
At $t \simeq 2.3$ \unit{Gyr}, an encounter with a $1.47\ \mathrm{M}_\odot$ MS star results in a companion exchange.
The NS--MS binary subsequently interacts with a triple system containing two MS stars of approximately $1.14\ \mathrm{M}_\odot$ each and a $0.63\ \mathrm{M}_\odot$ MS star.
The two $1.14\ \mathrm{M}_\odot$ stars merge, and by $t \simeq 2.76$ \unit{Gyr} the NS is bound to the resulting $2.29\ \mathrm{M}_\odot$ MS star.
A later encounter with a binary composed of $1.39$ and $1.01\ \mathrm{M}_\odot$ MS stars replaces this merger-product companion with the $1.01\ \mathrm{M}_\odot$ star, establishing the final NS--MS pair at $t_\mathrm{form} \simeq 2.93$ \unit{Gyr}.
The resulting binary has $P \sim 10^4$ days.
Subsequent dynamical perturbations modify its orbit, and it escapes from the cluster at $t \simeq 2.944$ \unit{Gyr} with $P \simeq 6300$ days and $e \simeq 0.69$.

At $t=1.6$ Gyr, the turn-off mass ($\sim 1.8\ \mathrm{M}_\odot$) becomes comparable to the mass of the NS ($m_\mathrm{NS} = 1.87\ \mathrm{M}_\odot$).
Once the BHs are depleted, this near-equality of masses means that the NS can be retained as a member of the binary in exchange encounters, rather than being the ejected body.
In the core, the NS exchanges companions with other stars.
All dynamical exchanges observed after $1.6$ \unit{Gyr} happen within or near the cluster core.
This case illustrates a formation channel in which NSs undergo repeated dynamical interactions in the core and finally form wide binaries such as Gaia NSs.
In this particular case, however, BH depletion and the epoch at which the turn-off mass falls below the NS mass occur at almost the same time, so we cannot separate their individual contributions from this single example.

\section{Discussion}
\subsection{Comparison with previous studies}
Previous $N$-body studies have reported that the dynamical formation efficiency of Gaia NSs is extremely low, in particular in simulations that follow only the early evolution of star clusters.
For example, \citet{Tanikawa+2024NS} found that the formation efficiency of Gaia NSs
in open clusters is lower than that of Gaia BHs by at least an order of magnitude.
Similarly, \citet{Fantoccoli+2025} investigated compact-object binaries in more massive star clusters, where long-lived BH subsystems dominate the cluster core dynamics.
In their simulations, NS binaries are strongly suppressed, because the cluster core is dominated by BHs until the end of the simulation and NSs have few opportunities to acquire companions there.

The apparent discrepancy with our results can be understood in terms of the evolutionary phase being considered.
Previous studies primarily examined the early ($\lesssim 1$ Gyr) evolution of star clusters, whereas our simulations extend to $7$ Gyr.
This is well after the depletion of BHs from the cluster core, and after the MS turn-off mass has decreased to a value comparable to the NS mass.
In this late-time regime, the strong encounters in the core are no longer monopolized by BHs, and NSs can form Gaia NS-like binaries through exchange interactions.
The formation rate remains low, but this channel is open at late times.
The low yields of Gaia NS-like binaries in earlier studies may therefore partly reflect the evolutionary phases sampled, rather than a general prohibition on their dynamical formation.
\citet{Tanikawa+2024NS} also noted that the main-sequence lifetime of a $1.4\ \mathrm{M}_\odot$ star is a few Gyr, longer than the lifetime of the open clusters they considered. 
Our models follow exactly this regime: the clusters survive to $7$ Gyr, and the Gaia-sensitive NS--MS binaries are assembled after the turn-off mass has fallen to a value comparable to the NS mass.

\subsection{Implications for the observed Gaia NS population}
\citet{Banerjee2026} recently investigated the formation of Gaia compact binaries in long-term simulations of $10^4$--$10^5\ \mathrm{M}_\odot$ cluster models.
That study showed that dynamically paired NS--MS binaries can appear once the cluster evolution is followed beyond the early BH-dominated phase.
Our results are consistent with this picture, but identify the underlying dynamical transition more specifically.
The NSs are already present in the core before BH depletion; what changes is that the strong encounters in the core cease to be monopolized by the BH subsystem.
At the same time, the NS--MS systems in our models are biased toward relatively long orbital periods and high eccentricities compared with the currently observed Gaia NS candidates.
This suggests that late-time cluster dynamics can produce NS--MS binaries in the Gaia-sensitive period range, but reproducing the full observed period--eccentricity distribution likely requires a broader exploration of cluster mass, radius, metallicity, velocity dispersion, and BH-retention history.

\subsection{Limitations and realism of the initial conditions}
The purpose of this paper is to investigate the dynamical formation of Gaia NSs under the assumption that at least a subset of the observed Gaia NS population originates from star clusters.
In particular, we aim to demonstrate that, in contrast to Gaia BHs, Gaia NSs are preferentially formed during a later evolutionary phase of star clusters, after the depletion of the BH subsystem.
Accordingly, this study is not intended to provide a comprehensive survey of formation efficiencies as a function of cluster mass, metallicity, or initial structural parameters, nor to model fully realistic YMCs.
Our simulations are controlled numerical experiments.
They isolate the late-time dynamical regime relevant for Gaia NS formation, and they clarify the physical conditions under which such systems can form.

The cluster models adopted in this study do not represent the full diversity of YMCs, but their initial masses and radii lie at the compact end of the observed cluster population.
Observed young and massive star clusters span a broad range of masses and radii, including systems with masses of order $10^4\ \mathrm{M}_\odot$ and characteristic half-mass radii below $1$ pc \citep{PortegiesZwart+2010, Krumholz+2019}.
For example, a $10^4\,\mathrm{M}_\odot$ cluster at a Galactocentric radius of $8$ kpc would be classified as a YMC.
If the ratio of the half-mass radius to the tidal radius ($r_\mathrm{h}/r_\mathrm{t}$) is the same, the characteristic dynamical timescales scale self-consistently.
Only the stellar evolution timescale does not scale in this way.

In this case, the corresponding half-mass radius is $\sim 0.38$ pc if we assume $r_\mathrm{t}\propto R_{\rm G}^{2/3}$, which is comparable to the characteristic half-mass radius of $\sim0.33$ pc expected for $10^4\ \mathrm{M}_\odot$ clusters from the mass--radius relation of \citet{MarksKroupa2012}.
Therefore, the initial densities explored in this work fall within the range expected for compact star clusters.
Thus, we expect that the qualitative separation of dynamical formation phases would remain valid in more realistic YMC models, although the absolute timescales may differ.
Clusters that survive for several Gyr and become BH-poor enter a similar dynamical regime, regardless of their early evolutionary history.

\section{Conclusions}
In this paper, we performed direct $N$-body simulations of $10^4\ \mathrm{M}_\odot$ star clusters, and investigated the dynamical formation of Gaia NS-like binaries after the depletion of the BH subsystem.

Our results support the interpretation that the BH subsystem suppresses the dynamical formation of Gaia NS-like binaries even when some NSs are already present in the core.
Across 20 realizations, we identify three dynamically formed NS--MS binaries that satisfy our adopted Gaia-sensitive criteria, all of which form after BH depletion.
Two short-period NS--MS binaries form and escape at early times, but they do not satisfy these criteria.
Our sample is small and is restricted to one class of compact $10^4\ \mathrm{M}_\odot$ clusters, and it does not provide a robust formation efficiency.

The NS--MS binaries formed in our models tend to have relatively long orbital periods and high eccentricities.
Late-time cluster dynamics therefore produce the wide and eccentric subset of Gaia NS-like systems.
The full observed population may require additional channels or different cluster environments.

A natural next step is to determine what sets the orbital-period scale of dynamically formed NS--MS binaries.
\citet{Banerjee2026} explored a broad grid of cluster models and showed that Gaia NS-like systems can be produced in long-term cluster simulations.
However, the connection between the cluster dynamical state and the final orbital periods of NS--MS binaries remains unclear.
In our models, the NS--MS systems are biased toward long periods.
Late-time assembly alone is therefore not sufficient to populate the short-period Gaia NS regime efficiently.
Future simulations should therefore explore cluster initial conditions and modeling choices, with particular emphasis on the velocity dispersion and the density evolution of the host cluster.
The cluster velocity dispersion sets the hard--soft boundary, and the core density and the remaining cluster lifetime set the number of subsequent hardening encounters.
The relation between the final orbital periods of NS--MS binaries and these cluster-scale quantities will show whether the observed short-period Gaia NS candidates require denser or more massive clusters than those studied here.

\begin{acknowledgments}
We thank the anonymous referee for careful reading and helpful suggestions that improved this paper.
R. I. acknowledges support from JST SPRING and SPRING Premium Program in Kobe University, Grant Number JPMJSP2148.
L.W. acknowledges the support from the National Natural Science Foundation of China through grants 12233013 and 12573041, the GuangDong Basic and Applied Basic Research Foundation (2026A1515012460), the High-level Youth Talent Project (Provincial Financial Allocation) through the grant 2023HYSPT0706, the Fundamental Research Funds for the Central Universities, Sun Yat-sen University (2025QNPY04).
A. T. was supported by Grants-in-Aid for Scientific Research, 24K07040 and 25K01035, from the Japan Society for the Promotion of Science.
Our numerical simulations were carried out on the Personal Computer Cluster (PPC) and the NC-E cluster operated by the Beijing Super Cloud Center (BSCC), China, URL: \url{http://www.blsc.cn/}.
Numerical analyses were carried out on the CPU cluster ``CPLAB'' in Kobe University, and the analysis servers at the CfCA of NAOJ.
We used ChatGPT and Claude to improve the language of the manuscript.
\end{acknowledgments}

\software{\textsc{petar} \citep{petar}, \textsc{sdar} \citep{sdar}, \textsc{fdps} \citep{Iwasawa+2016}, \textsc{mcluster} \citep{Kuepper+2011, Wang+2019}, \textsc{bse} \citep{Hurley+2000,Hurley+2002, Banerjee+2020}, \textsc{galpy} \citep{galpy}, \textsc{numpy} \citep{Harris+2020} and \textsc{matplotlib} \citep{Hunter2007}.
}

\appendix
\section{Evolution of individual realizations}
\label{appendix}
\begin{figure*}[h]
    \centering
    \includegraphics[width=\textwidth]{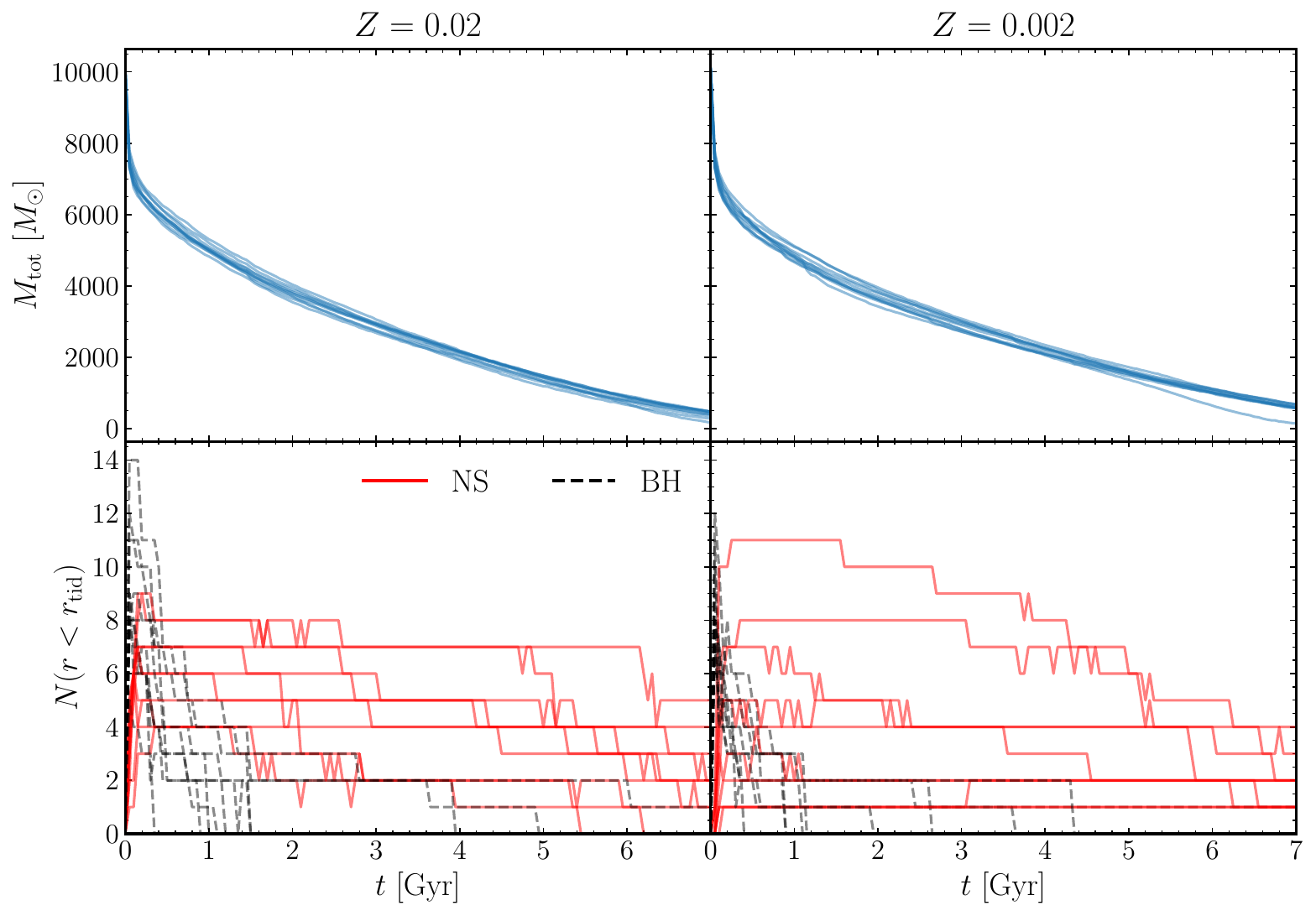}
    \caption{
    Evolution of the cluster mass and compact-object populations in the individual realizations.
    Each curve represents one realization.
    The upper panels show the stellar mass enclosed within the
    instantaneous tidal radius, while the lower panels show the
    numbers of BHs and NSs within the instantaneous tidal radius.
    }
    \label{fig:individual_runs}
\end{figure*}

Figure \ref{fig:individual_runs} shows the evolution presented in Figure \ref{fig:num_of_compact_objects} separately for each realization.
Although the BH-depletion times and the numbers of retained NSs vary among the realizations, the overall evolutionary trends are consistent with those shown by the ensemble averages in Figure \ref{fig:num_of_compact_objects}.
In most realizations, fewer than two BHs remain by the first $\sim1$--$2$ Gyr.
In several realizations, however, two BHs survive beyond $3$ Gyr.
Each of the three Gaia-sensitive NS--MS binaries listed in Table \ref{tab:ns_info} forms after BH depletion in its own realization.
The short-term fluctuations in the number of NSs have two origins: 
some NSs near the tidal radius cross it back and forth, and new NSs form through AIC at late times.

\bibliography{manuscript}

@ARTICLE{petar,
       author = {{Wang}, Long and {Iwasawa}, Masaki and {Nitadori}, Keigo and {Makino}, Junichiro},
        title = "{PETAR: a high-performance N-body code for modelling massive collisional stellar systems}",
      journal = {\mnras},
         year = 2020,
        month = sep,
       volume = {497},
       number = {1},
        pages = {536-555},
          doi = {10.1093/mnras/staa1915},
archivePrefix = {arXiv},
       eprint = {2006.16560},
 primaryClass = {astro-ph.IM},
       adsurl = {https://ui.adsabs.harvard.edu/abs/2020MNRAS.497..536W}
}

@Article{Numpy,
 title = {Array programming with {NumPy}},
 author = {Charles R. Harris and K. Jarrod Millman and St{'{e}}fan J. van der Walt and Ralf Gommers and Pauli Virtanen and David
                 Cournapeau and Eric Wieser and Julian Taylor and Sebastian
                 Berg and Nathaniel J. Smith and Robert Kern and Matti Picus
                 and Stephan Hoyer and Marten H. van Kerkwijk and Matthew
                 Brett and Allan Haldane and Jaime Fern{'{a}}ndez del
                 R{'{\i}}o and Mark Wiebe and Pearu Peterson and Pierre
                 G{'{e}}rard-Marchant and Kevin Sheppard and Tyler Reddy and
                 Warren Weckesser and Hameer Abbasi and Christoph Gohlke and
                 Travis E. Oliphant},
 year          = {2020},
 month         = sep,
 journal       = {Nature},
 volume        = {585},
 number        = {7825},
 pages         = {357--362},
 doi           = {10.1038/s41586-020-2649-2},
 publisher     = {Springer Science and Business Media {LLC}},
 url           = {https://doi.org/10.1038/s41586-020-2649-2}
}

@Article{Matplotlib,
  Author    = {Hunter, J. D.},
  Title     = {Matplotlib: A 2D graphics environment},
  Journal   = {Computing in Science \& Engineering},
  Volume    = {9},
  Number    = {3},
  Pages     = {90--95},
  publisher = {IEEE COMPUTER SOC},
  doi       = {10.1109/MCSE.2007.55},
  year      = 2007
}

@ARTICLE{Balbinot+2024,
       author = {{Balbinot}, E. and {Dodd}, E. and {Matsuno}, T. and {Lardo}, C. and {Helmi}, A. and {Panuzzo}, P. and {Mazeh}, T. and {Holl}, B. and {Caffau}, E. and {Jorissen}, A. and {Babusiaux}, C. and {Gavras}, P. and {Wyrzykowski}, {\L}. and {Eyer}, L. and {Leclerc}, N. and {Bombrun}, A. and {Mowlavi}, N. and {Seabroke}, G.~M. and {Cabrera-Ziri}, I. and {Callingham}, T.~M. and {Ruiz-Lara}, T. and {Starkenburg}, E.},
        title = "{The 33 M$_{{\ensuremath{\odot}}}$ black hole Gaia BH3 is part of the disrupted ED-2 star cluster}",
      journal = {\aap},
         year = 2024,
        month = jul,
       volume = {687},
          eid = {L3},
        pages = {L3},
          doi = {10.1051/0004-6361/202450425},
archivePrefix = {arXiv},
       eprint = {2404.11604},
 primaryClass = {astro-ph.GA},
       adsurl = {https://ui.adsabs.harvard.edu/abs/2024A&A...687L...3B}
}

@ARTICLE{GaiaBH3,
       author = {{Gaia Collaboration} and {Panuzzo}, P. and {Mazeh}, T. and {Arenou}, F. and {Holl}, B. and {Caffau}, E. and {Jorissen}, A. and {Babusiaux}, C. and {Gavras}, P. and {Sahlmann}, J. and {Bastian}, U. and {Wyrzykowski}, {\L}. and {Eyer}, L. and {Leclerc}, N. and {Bauchet}, N. and {Bombrun}, A. and {Mowlavi}, N. and {Seabroke}, G.~M. and {Teyssier}, D. and {Balbinot}, E. and {Helmi}, A. and {Brown}, A.~G.~A. and {Vallenari}, A. and {Prusti}, T. and {de Bruijne}, J.~H.~J. and {Barbier}, A. and {Biermann}, M. and {Creevey}, O.~L. and {Ducourant}, C. and {Evans}, D.~W. and {Guerra}, R. and {Hutton}, A. and {Jordi}, C. and {Klioner}, S.~A. and {Lammers}, U. and {Lindegren}, L. and {Luri}, X. and {Mignard}, F. and {Nicolas}, C. and {Randich}, S. and {Sartoretti}, P. and {Smiljanic}, R. and {Tanga}, P. and {Walton}, N.~A. and {Aerts}, C. and {Bailer-Jones}, C.~A.~L. and {Cropper}, M. and {Drimmel}, R. and {Jansen}, F. and {Katz}, D. and {Lattanzi}, M.~G. and {Soubiran}, C. and {Th{\'e}venin}, F. and {van Leeuwen}, F. and {Andrae}, R. and {Audard}, M. and {Bakker}, J. and {Blomme}, R. and {Casta{\~n}eda}, J. and {De Angeli}, F. and {Fabricius}, C. and {Fouesneau}, M. and {Fr{\'e}mat}, Y. and {Galluccio}, L. and {Guerrier}, A. and {Heiter}, U. and {Masana}, E. and {Messineo}, R. and {Nienartowicz}, K. and {Pailler}, F. and {Riclet}, F. and {Roux}, W. and {Sordo}, R. and {Gracia-Abril}, G. and {Portell}, J. and {Altmann}, M. and {Benson}, K. and {Berthier}, J. and {Burgess}, P.~W. and {Busonero}, D. and {Busso}, G. and {Cacciari}, C. and {C{\'a}novas}, H. and {Carrasco}, J.~M. and {Carry}, B. and {Cellino}, A. and {Cheek}, N. and {Clementini}, G. and {Damerdji}, Y. and {Davidson}, M. and {de Teodoro}, P. and {Delchambre}, L. and {Dell'Oro}, A. and {Fraile Garcia}, E. and {Garabato}, D. and {Garc{\'\i}a-Lario}, P. and {Haigron}, R. and {Hambly}, N.~C. and {Harrison}, D.~L. and {Hatzidimitriou}, D. and {Hern{\'a}ndez}, J. and {Hestroffer}, D. and {Hodgkin}, S.~T. and {Jamal}, S. and {Jevardat de Fombelle}, G. and {Jordan}, S. and {Krone-Martins}, A. and {Lanzafame}, A.~C. and {L{\"o}ffler}, W. and {Lorca}, A. and {Marchal}, O. and {Marrese}, P.~M. and {Moitinho}, A. and {Muinonen}, K. and {Nu{\~n}ez Campos}, M. and {Oreshina-Slezak}, I. and {Osborne}, P. and {Pancino}, E. and {Pauwels}, T. and {Recio-Blanco}, A. and {Riello}, M. and {Rimoldini}, L. and {Robin}, A.~C. and {Roegiers}, T. and {Sarro}, L.~M. and {Schultheis}, M. and {Smith}, M. and {Sozzetti}, A. and {Utrilla}, E. and {van Leeuwen}, M. and {Weingrill}, K. and {Abbas}, U. and {{\'A}brah{\'a}m}, P. and {Abreu Aramburu}, A. and {Ahmed}, S. and {Altavilla}, G. and {{\'A}lvarez}, M.~A. and {Anders}, F. and {Anderson}, R.~I. and {Anglada Varela}, E. and {Antoja}, T. and {Baig}, S. and {Baines}, D. and {Baker}, S.~G. and {Balaguer-N{\'u}{\~n}ez}, L. and {Balog}, Z. and {Barache}, C. and {Barros}, M. and {Barstow}, M.~A. and {Bartolom{\'e}}, S. and {Bashi}, D. and {Bassilana}, J. -L. and {Baudeau}, N. and {Becciani}, U. and {Bedin}, L.~R. and {Bellas-Velidis}, I. and {Bellazzini}, M. and {Beordo}, W. and {Bernet}, M. and {Bertolotto}, C. and {Bertone}, S. and {Bianchi}, L. and {Binnenfeld}, A. and {Blanco-Cuaresma}, S. and {Bland-Hawthorn}, J. and {Blazere}, A. and {Boch}, T. and {Bossini}, D. and {Bouquillon}, S. and {Bragaglia}, A. and {Braine}, J. and {Bratsolis}, E. and {Breedt}, E. and {Bressan}, A. and {Brouillet}, N. and {Brugaletta}, E. and {Bucciarelli}, B. and {Butkevich}, A.~G. and {Buzzi}, R. and {Camut}, A. and {Cancelliere}, R. and {Cantat-Gaudin}, T. and {Capilla Guilarte}, D. and {Carballo}, R. and {Carlucci}, T. and {Carnerero}, M.~I. and {Carretero}, J. and {Carton}, S. and {Casamiquela}, L. and {Casey}, A. and {Castellani}, M. and {Castro-Ginard}, A. and {Ceraj}, L. and {Cesare}, V. and {Charlot}, P. and {Chaudet}, C. and {Chemin}, L. and {Chiavassa}, A. and {Chornay}, N. and {Chosson}, D.},
        title = "{Discovery of a dormant 33 solar-mass black hole in pre-release Gaia astrometry}",
      journal = {\aap},
         year = 2024,
        month = jun,
       volume = {686},
          eid = {L2},
        pages = {L2},
          doi = {10.1051/0004-6361/202449763},
archivePrefix = {arXiv},
       eprint = {2404.10486},
 primaryClass = {astro-ph.GA},
       adsurl = {https://ui.adsabs.harvard.edu/abs/2024A&A...686L...2G}
}

@ARTICLE{Oshino+2011,
       author = {{Oshino}, Shoichi and {Funato}, Yoko and {Makino}, Junichiro},
        title = "{Particle-Particle Particle-Tree: A Direct-Tree Hybrid Scheme for Collisional N-Body Simulations}",
      journal = {\pasj},
         year = 2011,
        month = aug,
       volume = {63},
        pages = {881},
          doi = {10.1093/pasj/63.4.881},
archivePrefix = {arXiv},
       eprint = {1101.5504},
 primaryClass = {astro-ph.EP},
       adsurl = {https://ui.adsabs.harvard.edu/abs/2011PASJ...63..881O}
}

@ARTICLE{sdar,
       author = {{Wang}, Long and {Nitadori}, Keigo and {Makino}, Junichiro},
        title = "{A slow-down time-transformed symplectic integrator for solving the few-body problem}",
      journal = {\mnras},
         year = 2020,
        month = apr,
       volume = {493},
       number = {3},
        pages = {3398-3411},
          doi = {10.1093/mnras/staa480},
archivePrefix = {arXiv},
       eprint = {2002.07938},
 primaryClass = {astro-ph.EP},
       adsurl = {https://ui.adsabs.harvard.edu/abs/2020MNRAS.493.3398W}
}

@ARTICLE{BarnesHut86,
       author = {{Barnes}, Josh and {Hut}, Piet},
        title = "{A hierarchical O(N log N) force-calculation algorithm}",
      journal = {\nat},
         year = 1986,
        month = dec,
       volume = {324},
       number = {6096},
        pages = {446-449},
          doi = {10.1038/324446a0},
       adsurl = {https://ui.adsabs.harvard.edu/abs/1986Natur.324..446B}
}

@ARTICLE{Iwasawa+2016,
       author = {{Iwasawa}, Masaki and {Tanikawa}, Ataru and {Hosono}, Natsuki and {Nitadori}, Keigo and {Muranushi}, Takayuki and {Makino}, Junichiro},
        title = "{Implementation and performance of FDPS: a framework for developing parallel particle simulation codes}",
      journal = {\pasj},
         year = 2016,
        month = aug,
       volume = {68},
       number = {4},
          eid = {54},
        pages = {54},
          doi = {10.1093/pasj/psw053},
archivePrefix = {arXiv},
       eprint = {1601.03138},
 primaryClass = {astro-ph.IM},
       adsurl = {https://ui.adsabs.harvard.edu/abs/2016PASJ...68...54I}
}

@ARTICLE{Iwasawa+2020,
       author = {{Iwasawa}, Masaki and {Namekata}, Daisuke and {Nitadori}, Keigo and {Nomura}, Kentaro and {Wang}, Long and {Tsubouchi}, Miyuki and {Makino}, Junichiro},
        title = "{Accelerated FDPS: Algorithms to use accelerators with FDPS}",
      journal = {\pasj},
         year = 2020,
        month = feb,
       volume = {72},
       number = {1},
          eid = {13},
        pages = {13},
          doi = {10.1093/pasj/psz133},
archivePrefix = {arXiv},
       eprint = {1907.02290},
 primaryClass = {astro-ph.IM},
       adsurl = {https://ui.adsabs.harvard.edu/abs/2020PASJ...72...13I}
}

@ARTICLE{galpy,
       author = {{Bovy}, Jo},
        title = "{galpy: A python Library for Galactic Dynamics}",
      journal = {\apjs},
         year = 2015,
        month = feb,
       volume = {216},
       number = {2},
          eid = {29},
        pages = {29},
          doi = {10.1088/0067-0049/216/2/29},
archivePrefix = {arXiv},
       eprint = {1412.3451},
 primaryClass = {astro-ph.GA},
       adsurl = {https://ui.adsabs.harvard.edu/abs/2015ApJS..216...29B}
}

@ARTICLE{Hurley+2000,
       author = {{Hurley}, Jarrod R. and {Pols}, Onno R. and {Tout}, Christopher A.},
        title = "{Comprehensive analytic formulae for stellar evolution as a function of mass and metallicity}",
      journal = {\mnras},
         year = 2000,
        month = jul,
       volume = {315},
       number = {3},
        pages = {543-569},
          doi = {10.1046/j.1365-8711.2000.03426.x},
archivePrefix = {arXiv},
       eprint = {astro-ph/0001295},
 primaryClass = {astro-ph},
       adsurl = {https://ui.adsabs.harvard.edu/abs/2000MNRAS.315..543H}
}

@ARTICLE{Hurley+2002,
       author = {{Hurley}, Jarrod R. and {Tout}, Christopher A. and {Pols}, Onno R.},
        title = "{Evolution of binary stars and the effect of tides on binary populations}",
      journal = {\mnras},
         year = 2002,
        month = feb,
       volume = {329},
       number = {4},
        pages = {897-928},
          doi = {10.1046/j.1365-8711.2002.05038.x},
archivePrefix = {arXiv},
       eprint = {astro-ph/0201220},
 primaryClass = {astro-ph},
       adsurl = {https://ui.adsabs.harvard.edu/abs/2002MNRAS.329..897H}
}

@ARTICLE{Banerjee+2020,
       author = {{Banerjee}, S. and {Belczynski}, K. and {Fryer}, C.~L. and {Berczik}, P. and {Hurley}, J.~R. and {Spurzem}, R. and {Wang}, L.},
        title = "{BSE versus StarTrack: Implementations of new wind, remnant-formation, and natal-kick schemes in NBODY7 and their astrophysical consequences}",
      journal = {\aap},
         year = 2020,
        month = jul,
       volume = {639},
          eid = {A41},
        pages = {A41},
          doi = {10.1051/0004-6361/201935332},
archivePrefix = {arXiv},
       eprint = {1902.07718},
 primaryClass = {astro-ph.SR},
       adsurl = {https://ui.adsabs.harvard.edu/abs/2020A&A...639A..41B}
}

@ARTICLE{MarinPina+2024,
       author = {{Mar{\'\i}n Pina}, Daniel and {Rastello}, Sara and {Gieles}, Mark and {Kremer}, Kyle and {Fitzgerald}, Laura and {Rando Forastier}, Bruno},
        title = "{Dynamical formation of Gaia BH3 in the progenitor globular cluster of the ED-2 stream}",
      journal = {\aap},
         year = 2024,
        month = aug,
       volume = {688},
          eid = {L2},
        pages = {L2},
          doi = {10.1051/0004-6361/202450460},
archivePrefix = {arXiv},
       eprint = {2404.13036},
 primaryClass = {astro-ph.GA},
       adsurl = {https://ui.adsabs.harvard.edu/abs/2024A&A...688L...2M}
}

@ARTICLE{El-Badry+2024NS21,
       author = {{El-Badry}, Kareem and {Rix}, Hans-Walter and {Latham}, David W. and {Shahaf}, Sahar and {Mazeh}, Tsevi and {Bieryla}, Allyson and {Buchhave}, Lars A. and {Andrae}, Ren{\'e} and {Yamaguchi}, Natsuko and {Isaacson}, Howard and {Howard}, Andrew W. and {Savino}, Alessandro and {Ilyin}, Ilya V.},
        title = "{A population of neutron star candidates in wide orbits from Gaia astrometry}",
      journal = {The Open Journal of Astrophysics},
         year = 2024,
        month = jul,
       volume = {7},
          eid = {58},
        pages = {58},
          doi = {10.33232/001c.121261},
archivePrefix = {arXiv},
       eprint = {2405.00089},
 primaryClass = {astro-ph.SR},
       adsurl = {https://ui.adsabs.harvard.edu/abs/2024OJAp....7E..58E}
}

@ARTICLE{El-Badry+2024NS1,
       author = {{El-Badry}, Kareem and {Simon}, Joshua D. and {Reggiani}, Henrique and {Rix}, Hans-Walter and {Latham}, David W. and {Bieryla}, Allyson and {Buchhave}, Lars A. and {Shahaf}, Sahar and {Mazeh}, Tsevi and {Chakrabarti}, Sukanya and {Guhathakurta}, Puragra and {Ilyin}, Ilya V. and {Tauris}, Thomas M.},
        title = "{A 1.9 solar-mass neutron star candidate in a 2-year orbit}",
      journal = {The Open Journal of Astrophysics},
         year = 2024,
        month = apr,
       volume = {7},
          eid = {27},
        pages = {27},
          doi = {10.33232/001c.116675},
archivePrefix = {arXiv},
       eprint = {2402.06722},
 primaryClass = {astro-ph.SR},
       adsurl = {https://ui.adsabs.harvard.edu/abs/2024OJAp....7E..27E}
}

@ARTICLE{El-Badry+2023BH2,
       author = {{El-Badry}, Kareem and {Rix}, Hans-Walter and {Cendes}, Yvette and {Rodriguez}, Antonio C. and {Conroy}, Charlie and {Quataert}, Eliot and {Hawkins}, Keith and {Zari}, Eleonora and {Hobson}, Melissa and {Breivik}, Katelyn and {Rau}, Arne and {Berger}, Edo and {Shahaf}, Sahar and {Seeburger}, Rhys and {Burdge}, Kevin B. and {Latham}, David W. and {Buchhave}, Lars A. and {Bieryla}, Allyson and {Bashi}, Dolev and {Mazeh}, Tsevi and {Faigler}, Simchon},
        title = "{A red giant orbiting a black hole}",
      journal = {\mnras},
         year = 2023,
        month = may,
       volume = {521},
       number = {3},
        pages = {4323-4348},
          doi = {10.1093/mnras/stad799},
archivePrefix = {arXiv},
       eprint = {2302.07880},
 primaryClass = {astro-ph.SR},
       adsurl = {https://ui.adsabs.harvard.edu/abs/2023MNRAS.521.4323E}
}

@ARTICLE{El-Badry+2023BH1,
       author = {{El-Badry}, Kareem and {Rix}, Hans-Walter and {Quataert}, Eliot and {Howard}, Andrew W. and {Isaacson}, Howard and {Fuller}, Jim and {Hawkins}, Keith and {Breivik}, Katelyn and {Wong}, Kaze W.~K. and {Rodriguez}, Antonio C. and {Conroy}, Charlie and {Shahaf}, Sahar and {Mazeh}, Tsevi and {Arenou}, Fr{\'e}d{\'e}ric and {Burdge}, Kevin B. and {Bashi}, Dolev and {Faigler}, Simchon and {Weisz}, Daniel R. and {Seeburger}, Rhys and {Almada Monter}, Silvia and {Wojno}, Jennifer},
        title = "{A Sun-like star orbiting a black hole}",
      journal = {\mnras},
         year = 2023,
        month = jan,
       volume = {518},
       number = {1},
        pages = {1057-1085},
          doi = {10.1093/mnras/stac3140},
archivePrefix = {arXiv},
       eprint = {2209.06833},
 primaryClass = {astro-ph.SR},
       adsurl = {https://ui.adsabs.harvard.edu/abs/2023MNRAS.518.1057E}
}

@ARTICLE{Chakrabarti+2023,
       author = {{Chakrabarti}, Sukanya and {Simon}, Joshua D. and {Craig}, Peter A. and {Reggiani}, Henrique and {Brandt}, Timothy D. and {Guhathakurta}, Puragra and {Dalba}, Paul A. and {Kirby}, Evan N. and {Chang}, Philip and {Hey}, Daniel R. and {Savino}, Alessandro and {Geha}, Marla and {Thompson}, Ian B.},
        title = "{A Noninteracting Galactic Black Hole Candidate in a Binary System with a Main-sequence Star}",
      journal = {\aj},
         year = 2023,
        month = jul,
       volume = {166},
       number = {1},
          eid = {6},
        pages = {6},
          doi = {10.3847/1538-3881/accf21},
archivePrefix = {arXiv},
       eprint = {2210.05003},
 primaryClass = {astro-ph.GA},
       adsurl = {https://ui.adsabs.harvard.edu/abs/2023AJ....166....6C}
}

@ARTICLE{Kremer+2020,
       author = {{Kremer}, Kyle and {Ye}, Claire S. and {Rui}, Nicholas Z. and {Weatherford}, Newlin C. and {Chatterjee}, Sourav and {Fragione}, Giacomo and {Rodriguez}, Carl L. and {Spera}, Mario and {Rasio}, Frederic A.},
        title = "{Modeling Dense Star Clusters in the Milky Way and Beyond with the CMC Cluster Catalog}",
      journal = {\apjs},
         year = 2020,
        month = apr,
       volume = {247},
       number = {2},
          eid = {48},
        pages = {48},
          doi = {10.3847/1538-4365/ab7919},
archivePrefix = {arXiv},
       eprint = {1911.00018},
 primaryClass = {astro-ph.HE},
       adsurl = {https://ui.adsabs.harvard.edu/abs/2020ApJS..247...48K}
}

@ARTICLE{Tanikawa+2024,
       author = {{Tanikawa}, Ataru and {Cary}, Savannah and {Shikauchi}, Minori and {Wang}, Long and {Fujii}, Michiko S.},
        title = "{Compact binary formation in open star clusters - I. High formation efficiency of Gaia BHs and their multiplicities}",
      journal = {\mnras},
         year = 2024,
        month = jan,
       volume = {527},
       number = {2},
        pages = {4031-4039},
          doi = {10.1093/mnras/stad3294},
archivePrefix = {arXiv},
       eprint = {2303.05743},
 primaryClass = {astro-ph.GA},
       adsurl = {https://ui.adsabs.harvard.edu/abs/2024MNRAS.527.4031T}
}

@ARTICLE{Tanikawa+2024NS,
       author = {{Tanikawa}, Ataru and {Wang}, Long and {Fujii}, Michiko S.},
        title = "{Compact Binary Formation in Open Star Clusters II: Difficulty of Gaia NS formation in low-mass star clusters}",
      journal = {The Open Journal of Astrophysics},
         year = 2024,
        month = may,
       volume = {7},
          eid = {39},
        pages = {39},
          doi = {10.33232/001c.117684},
archivePrefix = {arXiv},
       eprint = {2404.01731},
 primaryClass = {astro-ph.SR},
       adsurl = {https://ui.adsabs.harvard.edu/abs/2024OJAp....7E..39T}
}

@ARTICLE{Kuepper+2011,
       author = {{K{\"u}pper}, Andreas H.~W. and {Maschberger}, Thomas and {Kroupa}, Pavel and {Baumgardt}, Holger},
        title = "{Mass segregation and fractal substructure in young massive clusters - I. The McLuster code and method calibration}",
      journal = {\mnras},
         year = 2011,
        month = nov,
       volume = {417},
       number = {3},
        pages = {2300-2317},
          doi = {10.1111/j.1365-2966.2011.19412.x},
archivePrefix = {arXiv},
       eprint = {1107.2395},
 primaryClass = {astro-ph.GA},
       adsurl = {https://ui.adsabs.harvard.edu/abs/2011MNRAS.417.2300K}
}

@ARTICLE{MakinoAarseth92,
       author = {{Makino}, Junichiro and {Aarseth}, Sverre J.},
        title = "{On a Hermite Integrator with Ahmad-Cohen Scheme for Gravitational Many-Body Problems}",
      journal = {\pasj},
         year = 1992,
        month = apr,
       volume = {44},
        pages = {141-151},
       adsurl = {https://ui.adsabs.harvard.edu/abs/1992PASJ...44..141M}
}

@ARTICLE{Fantoccoli+2025,
       author = {{Fantoccoli}, Federico and {Barber}, Jordan and {Dosopoulou}, Fani and {Chattopadhyay}, Debatri and {Antonini}, Fabio},
        title = "{Properties of black hole-star binaries formed in N-body simulations of massive star clusters: implications for Gaia black holes}",
      journal = {\mnras},
         year = 2025,
        month = mar,
       volume = {538},
       number = {1},
        pages = {243-257},
          doi = {10.1093/mnras/staf303},
archivePrefix = {arXiv},
       eprint = {2410.17323},
 primaryClass = {astro-ph.GA},
       adsurl = {https://ui.adsabs.harvard.edu/abs/2025MNRAS.538..243F}
}

@ARTICLE{Gilkil&Mazeh2024,
       author = {{Gilkis}, A. and {Mazeh}, T.},
        title = "{Gaia BH1 and BH2 - evolutionary models with overshooting of the black hole progenitors within the present-day binary separation}",
      journal = {\mnras},
         year = 2024,
        month = nov,
       volume = {535},
       number = {1},
        pages = {L44-L48},
          doi = {10.1093/mnrasl/slae091},
archivePrefix = {arXiv},
       eprint = {2409.15899},
 primaryClass = {astro-ph.SR},
       adsurl = {https://ui.adsabs.harvard.edu/abs/2024MNRAS.535L..44G}
}

@ARTICLE{Fryer+2012,
       author = {{Fryer}, Chris L. and {Belczynski}, Krzysztof and {Wiktorowicz}, Grzegorz and {Dominik}, Michal and {Kalogera}, Vicky and {Holz}, Daniel E.},
        title = "{Compact Remnant Mass Function: Dependence on the Explosion Mechanism and Metallicity}",
      journal = {\apj},
         year = 2012,
        month = apr,
       volume = {749},
       number = {1},
          eid = {91},
        pages = {91},
          doi = {10.1088/0004-637X/749/1/91},
archivePrefix = {arXiv},
       eprint = {1110.1726},
 primaryClass = {astro-ph.SR},
       adsurl = {https://ui.adsabs.harvard.edu/abs/2012ApJ...749...91F}
}

@ARTICLE{Kroupa2001,
       author = {{Kroupa}, Pavel},
        title = "{On the variation of the initial mass function}",
      journal = {\mnras},
         year = 2001,
        month = apr,
       volume = {322},
       number = {2},
        pages = {231-246},
          doi = {10.1046/j.1365-8711.2001.04022.x},
archivePrefix = {arXiv},
       eprint = {astro-ph/0009005},
 primaryClass = {astro-ph},
       adsurl = {https://ui.adsabs.harvard.edu/abs/2001MNRAS.322..231K}
}

@ARTICLE{Plummer11,
       author = {{Plummer}, H.~C.},
        title = "{On the problem of distribution in globular star clusters}",
      journal = {\mnras},
         year = 1911,
        month = mar,
       volume = {71},
        pages = {460-470},
          doi = {10.1093/mnras/71.5.460},
       adsurl = {https://ui.adsabs.harvard.edu/abs/1911MNRAS..71..460P}
}

@ARTICLE{Wang+2019,
       author = {{Wang}, Long and {Kroupa}, Pavel and {Jerabkova}, Tereza},
        title = "{Complete ejection of OB stars from very young star clusters and the formation of multiple populations}",
      journal = {\mnras},
         year = 2019,
        month = apr,
       volume = {484},
       number = {2},
        pages = {1843-1851},
          doi = {10.1093/mnras/sty2232},
archivePrefix = {arXiv},
       eprint = {1810.07697},
 primaryClass = {astro-ph.SR},
       adsurl = {https://ui.adsabs.harvard.edu/abs/2019MNRAS.484.1843W}
}

@ARTICLE{Sana+2012,
       author = {{Sana}, H. and {de Mink}, S.~E. and {de Koter}, A. and {Langer}, N. and {Evans}, C.~J. and {Gieles}, M. and {Gosset}, E. and {Izzard}, R.~G. and {Le Bouquin}, J.-B. and {Schneider}, F.~R.~N.},
        title = "{Binary Interaction Dominates the Evolution of Massive Stars}",
      journal = {Science},
         year = 2012,
        month = jul,
       volume = {337},
       number = {6093},
        pages = {444},
          doi = {10.1126/science.1223344},
archivePrefix = {arXiv},
       eprint = {1207.6397},
 primaryClass = {astro-ph.SR},
       adsurl = {https://ui.adsabs.harvard.edu/abs/2012Sci...337..444S}
}

@ARTICLE{Kroupa95a,
       author = {{Kroupa}, Pavel},
        title = "{Inverse dynamical population synthesis and star formation}",
      journal = {\mnras},
         year = 1995,
        month = dec,
       volume = {277},
        pages = {1491},
          doi = {10.1093/mnras/277.4.1491},
archivePrefix = {arXiv},
       eprint = {astro-ph/9508117},
 primaryClass = {astro-ph},
       adsurl = {https://ui.adsabs.harvard.edu/abs/1995MNRAS.277.1491K}
}

@ARTICLE{Kroupa95b,
       author = {{Kroupa}, Pavel},
        title = "{The dynamical properties of stellar systems in the Galactic disc}",
      journal = {\mnras},
         year = 1995,
        month = dec,
       volume = {277},
        pages = {1507},
          doi = {10.1093/mnras/277.4.1507},
archivePrefix = {arXiv},
       eprint = {astro-ph/9508084},
 primaryClass = {astro-ph},
       adsurl = {https://ui.adsabs.harvard.edu/abs/1995MNRAS.277.1507K}
}

@ARTICLE{Belloni+2017,
       author = {{Belloni}, Diogo and {Askar}, Abbas and {Giersz}, Mirek and {Kroupa}, Pavel and {Rocha-Pinto}, Helio J.},
        title = "{On the initial binary population for star cluster simulations}",
      journal = {\mnras},
         year = 2017,
        month = nov,
       volume = {471},
       number = {3},
        pages = {2812-2828},
          doi = {10.1093/mnras/stx1763},
archivePrefix = {arXiv},
       eprint = {1707.04271},
 primaryClass = {astro-ph.GA},
       adsurl = {https://ui.adsabs.harvard.edu/abs/2017MNRAS.471.2812B}
}

@ARTICLE{Webbink84,
       author = {{Webbink}, R.~F.},
        title = "{Double white dwarfs as progenitors of R Coronae Borealis stars and type I supernovae.}",
      journal = {\apj},
         year = 1984,
        month = feb,
       volume = {277},
        pages = {355-360},
          doi = {10.1086/161701},
       adsurl = {https://ui.adsabs.harvard.edu/abs/1984ApJ...277..355W}
}

@ARTICLE{Baron+87,
       author = {{Baron}, E. and {Cooperstein}, J. and {Kahana}, S. and {Nomoto}, K.},
        title = "{Collapsing White Dwarfs}",
      journal = {\apj},
         year = 1987,
        month = sep,
       volume = {320},
        pages = {304},
          doi = {10.1086/165542},
       adsurl = {https://ui.adsabs.harvard.edu/abs/1987ApJ...320..304B}
}

@ARTICLE{NomotoKondo91,
       author = {{Nomoto}, Ken'ichi and {Kondo}, Yoji},
        title = "{Conditions for Accretion-induced Collapse of White Dwarfs}",
      journal = {\apjl},
         year = 1991,
        month = jan,
       volume = {367},
        pages = {L19},
          doi = {10.1086/185922},
       adsurl = {https://ui.adsabs.harvard.edu/abs/1991ApJ...367L..19N}
}

@ARTICLE{WoosleyBaron92,
       author = {{Woosley}, S.~E. and {Baron}, E.},
        title = "{The Collapse of White Dwarfs to Neutron Stars}",
      journal = {\apj},
         year = 1992,
        month = may,
       volume = {391},
        pages = {228},
          doi = {10.1086/171338},
       adsurl = {https://ui.adsabs.harvard.edu/abs/1992ApJ...391..228W}
}

@ARTICLE{Fryer+99,
       author = {{Fryer}, Chris and {Benz}, Willy and {Herant}, Marc and {Colgate}, Stirling A.},
        title = "{What Can the Accretion-induced Collapse of White Dwarfs Really Explain?}",
      journal = {\apj},
         year = 1999,
        month = may,
       volume = {516},
       number = {2},
        pages = {892-899},
          doi = {10.1086/307119},
archivePrefix = {arXiv},
       eprint = {astro-ph/9812058},
 primaryClass = {astro-ph},
       adsurl = {https://ui.adsabs.harvard.edu/abs/1999ApJ...516..892F}
}

@ARTICLE{Dessart+2006,
       author = {{Dessart}, L. and {Burrows}, A. and {Ott}, C.~D. and {Livne}, E. and {Yoon}, S.-C. and {Langer}, N.},
        title = "{Multidimensional Simulations of the Accretion-induced Collapse of White Dwarfs to Neutron Stars}",
      journal = {\apj},
         year = 2006,
        month = jun,
       volume = {644},
       number = {2},
        pages = {1063-1084},
          doi = {10.1086/503626},
archivePrefix = {arXiv},
       eprint = {astro-ph/0601603},
 primaryClass = {astro-ph},
       adsurl = {https://ui.adsabs.harvard.edu/abs/2006ApJ...644.1063D}
}

@ARTICLE{Abdikamalov+2010,
       author = {{Abdikamalov}, E.~B. and {Ott}, C.~D. and {Rezzolla}, L. and {Dessart}, L. and {Dimmelmeier}, H. and {Marek}, A. and {Janka}, H.-T.},
        title = "{Axisymmetric general relativistic simulations of the accretion-induced collapse of white dwarfs}",
      journal = {\prd},
         year = 2010,
        month = feb,
       volume = {81},
       number = {4},
          eid = {044012},
        pages = {044012},
          doi = {10.1103/PhysRevD.81.044012},
archivePrefix = {arXiv},
       eprint = {0910.2703},
 primaryClass = {astro-ph.HE},
       adsurl = {https://ui.adsabs.harvard.edu/abs/2010PhRvD..81d4012A}
}

@ARTICLE{LongoMicchi+2023,
       author = {{Longo Micchi}, Lu{\'\i}s Felipe and {Radice}, David and {Chirenti}, Cecilia},
        title = "{Multimessenger emission from the accretion-induced collapse of white dwarfs}",
      journal = {\mnras},
         year = 2023,
        month = nov,
       volume = {525},
       number = {4},
        pages = {6359-6376},
          doi = {10.1093/mnras/stad2420},
archivePrefix = {arXiv},
       eprint = {2306.04711},
 primaryClass = {astro-ph.HE},
       adsurl = {https://ui.adsabs.harvard.edu/abs/2023MNRAS.525.6359L}
}

@ARTICLE{Mori+2025,
       author = {{Mori}, Masamitsu and {Sawada}, Ryo and {Suwa}, Yudai and {Tanikawa}, Ataru and {Kashiyama}, Kazumi and {Murase}, Kohta},
        title = "{Gravitational collapse of white dwarfs to neutron stars: From initial conditions to explosions with neutrino-radiation hydrodynamics simulations}",
      journal = {\pasj},
         year = 2025,
        month = feb,
       volume = {77},
       number = {1},
        pages = {127-138},
          doi = {10.1093/pasj/psae104},
archivePrefix = {arXiv},
       eprint = {2306.17381},
 primaryClass = {astro-ph.HE},
       adsurl = {https://ui.adsabs.harvard.edu/abs/2025PASJ...77..127M}
}

@ARTICLE{Podsiadlowski+2004,
       author = {{Podsiadlowski}, Ph. and {Langer}, N. and {Poelarends}, A.~J.~T. and {Rappaport}, S. and {Heger}, A. and {Pfahl}, E.},
        title = "{The Effects of Binary Evolution on the Dynamics of Core Collapse and Neutron Star Kicks}",
      journal = {\apj},
         year = 2004,
        month = sep,
       volume = {612},
       number = {2},
        pages = {1044-1051},
          doi = {10.1086/421713},
archivePrefix = {arXiv},
       eprint = {astro-ph/0309588},
 primaryClass = {astro-ph},
       adsurl = {https://ui.adsabs.harvard.edu/abs/2004ApJ...612.1044P}
}

@ARTICLE{GessnerJanka2018,
       author = {{Gessner}, Alexandra and {Janka}, Hans-Thomas},
        title = "{Hydrodynamical Neutron-star Kicks in Electron-capture Supernovae and Implications for the CRAB Supernova}",
      journal = {\apj},
         year = 2018,
        month = sep,
       volume = {865},
       number = {1},
          eid = {61},
        pages = {61},
          doi = {10.3847/1538-4357/aadbae},
archivePrefix = {arXiv},
       eprint = {1802.05274},
 primaryClass = {astro-ph.HE},
       adsurl = {https://ui.adsabs.harvard.edu/abs/2018ApJ...865...61G}
}

@ARTICLE{Wang2018,
       author = {{Wang}, Bo},
        title = "{The single-degenerate model for the progenitors of accretion-induced collapse events}",
      journal = {\mnras},
         year = 2018,
        month = nov,
       volume = {481},
       number = {1},
        pages = {439-446},
          doi = {10.1093/mnras/sty2278},
archivePrefix = {arXiv},
       eprint = {1808.05992},
 primaryClass = {astro-ph.SR},
       adsurl = {https://ui.adsabs.harvard.edu/abs/2018MNRAS.481..439W}
}

@ARTICLE{WangLiu2020,
       author = {{Wang}, Bo and {Liu}, Dongdong},
        title = "{The formation of neutron star systems through accretion-induced collapse in white-dwarf binaries}",
      journal = {Research in Astronomy and Astrophysics},
         year = 2020,
        month = sep,
       volume = {20},
       number = {9},
          eid = {135},
        pages = {135},
          doi = {10.1088/1674-4527/20/9/135},
archivePrefix = {arXiv},
       eprint = {2005.01880},
 primaryClass = {astro-ph.SR},
       adsurl = {https://ui.adsabs.harvard.edu/abs/2020RAA....20..135W}
}

@ARTICLE{El-Badry2024,
       author = {{El-Badry}, Kareem},
        title = "{Gaia's binary star renaissance}",
      journal = {\nar},
         year = 2024,
        month = jun,
       volume = {98},
          eid = {101694},
        pages = {101694},
          doi = {10.1016/j.newar.2024.101694},
archivePrefix = {arXiv},
       eprint = {2403.12146},
 primaryClass = {astro-ph.SR},
       adsurl = {https://ui.adsabs.harvard.edu/abs/2024NewAR..9801694E}
}

@ARTICLE{Oh+2015,
       author = {{Oh}, Seungkyung and {Kroupa}, Pavel and {Pflamm-Altenburg}, Jan},
        title = "{Dependency of Dynamical Ejections of O Stars on the Masses of Very Young Star Clusters}",
      journal = {\apj},
         year = 2015,
        month = jun,
       volume = {805},
       number = {2},
          eid = {92},
        pages = {92},
          doi = {10.1088/0004-637X/805/2/92},
archivePrefix = {arXiv},
       eprint = {1503.08827},
 primaryClass = {astro-ph.GA},
       adsurl = {https://ui.adsabs.harvard.edu/abs/2015ApJ...805...92O}
}

@ARTICLE{Hobbs+2005,
       author = {{Hobbs}, G. and {Lorimer}, D.~R. and {Lyne}, A.~G. and {Kramer}, M.},
        title = "{A statistical study of 233 pulsar proper motions}",
      journal = {\mnras},
         year = 2005,
        month = jul,
       volume = {360},
       number = {3},
        pages = {974-992},
          doi = {10.1111/j.1365-2966.2005.09087.x},
archivePrefix = {arXiv},
       eprint = {astro-ph/0504584},
 primaryClass = {astro-ph},
       adsurl = {https://ui.adsabs.harvard.edu/abs/2005MNRAS.360..974H}
}

@ARTICLE{BreenHeggie2013,
       author = {{Breen}, Philip G. and {Heggie}, Douglas C.},
        title = "{Dynamical evolution of black hole subsystems in idealized star clusters}",
      journal = {\mnras},
         year = 2013,
        month = jul,
       volume = {432},
       number = {4},
        pages = {2779-2797},
          doi = {10.1093/mnras/stt628},
archivePrefix = {arXiv},
       eprint = {1304.3401},
 primaryClass = {astro-ph.GA},
       adsurl = {https://ui.adsabs.harvard.edu/abs/2013MNRAS.432.2779B}
}

@ARTICLE{Spitzer69,
       author = {{Spitzer}, Jr., Lyman},
        title = "{Equipartition and the Formation of Compact Nuclei in Spherical Stellar Systems}",
      journal = {\apjl},
         year = 1969,
        month = dec,
       volume = {158},
        pages = {L139},
          doi = {10.1086/180451},
       adsurl = {https://ui.adsabs.harvard.edu/abs/1969ApJ...158L.139S}
}

@ARTICLE{Wang2020,
       author = {{Wang}, Long},
        title = "{The survival of star clusters with black hole subsystems}",
      journal = {\mnras},
         year = 2020,
        month = jan,
       volume = {491},
       number = {2},
        pages = {2413-2423},
          doi = {10.1093/mnras/stz3179},
archivePrefix = {arXiv},
       eprint = {1911.05077},
 primaryClass = {astro-ph.GA},
       adsurl = {https://ui.adsabs.harvard.edu/abs/2020MNRAS.491.2413W}
}

@ARTICLE{Wang2022,
       author = {{Wang}, Long and {Tanikawa}, Ataru and {Fujii}, Michiko S.},
        title = "{The impact of primordial binary on the dynamical evolution of intermediate massive star clusters}",
      journal = {\mnras},
         year = 2022,
        month = feb,
       volume = {509},
       number = {4},
        pages = {4713-4722},
          doi = {10.1093/mnras/stab3255},
archivePrefix = {arXiv},
       eprint = {2110.15555},
 primaryClass = {astro-ph.GA},
       adsurl = {https://ui.adsabs.harvard.edu/abs/2022MNRAS.509.4713W}
}

@ARTICLE{Casertano&Hut85,
       author = {{Casertano}, S. and {Hut}, P.},
        title = "{Core radius and density measurements in N-body experiments Connections with theoretical and observational definitions}",
      journal = {\apj},
         year = 1985,
        month = nov,
       volume = {298},
        pages = {80-94},
          doi = {10.1086/163589},
       adsurl = {https://ui.adsabs.harvard.edu/abs/1985ApJ...298...80C}
}

@ARTICLE{MarksKroupa2012,
       author = {{Marks}, M. and {Kroupa}, P.},
        title = "{Inverse dynamical population synthesis. Constraining the initial conditions of young stellar clusters by studying their binary populations}",
      journal = {\aap},
         year = 2012,
        month = jul,
       volume = {543},
          eid = {A8},
        pages = {A8},
          doi = {10.1051/0004-6361/201118231},
archivePrefix = {arXiv},
       eprint = {1205.1508},
 primaryClass = {astro-ph.GA},
       adsurl = {https://ui.adsabs.harvard.edu/abs/2012A&A...543A...8M}
}

@ARTICLE{Tanikawa+2023GaiaBH,
       author = {{Tanikawa}, Ataru and {Hattori}, Kohei and {Kawanaka}, Norita and {Kinugawa}, Tomoya and {Shikauchi}, Minori and {Tsuna}, Daichi},
        title = "{Search for a Black Hole Binary in Gaia DR3 Astrometric Binary Stars with Spectroscopic Data}",
      journal = {\apj},
         year = 2023,
        month = apr,
       volume = {946},
       number = {2},
          eid = {79},
        pages = {79},
          doi = {10.3847/1538-4357/acbf36},
archivePrefix = {arXiv},
       eprint = {2209.05632},
 primaryClass = {astro-ph.SR},
       adsurl = {https://ui.adsabs.harvard.edu/abs/2023ApJ...946...79T}
}

@ARTICLE{Hirai+2024,
       author = {{Hirai}, Ryosuke and {Podsiadlowski}, Philipp and {Heger}, Alexander and {Nagakura}, Hiroki},
        title = "{Neutron Star Kicks plus Rockets as a Mechanism for Forming Wide Low-eccentricity Neutron Star Binaries}",
      journal = {\apjl},
         year = 2024,
        month = sep,
       volume = {972},
       number = {1},
          eid = {L18},
        pages = {L18},
          doi = {10.3847/2041-8213/ad6e77},
archivePrefix = {arXiv},
       eprint = {2407.20967},
 primaryClass = {astro-ph.HE},
       adsurl = {https://ui.adsabs.harvard.edu/abs/2024ApJ...972L..18H}
}

@ARTICLE{Verbunt+2017,
       author = {{Verbunt}, Frank and {Igoshev}, Andrei and {Cator}, Eric},
        title = "{The observed velocity distribution of young pulsars}",
      journal = {\aap},
         year = 2017,
        month = dec,
       volume = {608},
          eid = {A57},
        pages = {A57},
          doi = {10.1051/0004-6361/201731518},
archivePrefix = {arXiv},
       eprint = {1708.08281},
 primaryClass = {astro-ph.HE},
       adsurl = {https://ui.adsabs.harvard.edu/abs/2017A&A...608A..57V}
}

@ARTICLE{Banerjee2018,
       author = {{Banerjee}, Sambaran},
        title = "{Stellar-mass black holes in young massive and open stellar clusters and their role in gravitational-wave generation - II}",
      journal = {\mnras},
         year = 2018,
        month = jan,
       volume = {473},
       number = {1},
        pages = {909-926},
          doi = {10.1093/mnras/stx2347},
archivePrefix = {arXiv},
       eprint = {1707.00922},
 primaryClass = {astro-ph.HE},
       adsurl = {https://ui.adsabs.harvard.edu/abs/2018MNRAS.473..909B}
}

@ARTICLE{Fragione+2018,
       author = {{Fragione}, Giacomo and {Pavl{\'\i}k}, V{\'a}clav and {Banerjee}, Sambaran},
        title = "{Neutron stars and millisecond pulsars in star clusters: implications for the diffuse gamma-radiation from the Galactic Centre}",
      journal = {\mnras},
         year = 2018,
        month = nov,
       volume = {480},
       number = {4},
        pages = {4955-4962},
          doi = {10.1093/mnras/sty2234},
archivePrefix = {arXiv},
       eprint = {1804.04856},
 primaryClass = {astro-ph.GA},
       adsurl = {https://ui.adsabs.harvard.edu/abs/2018MNRAS.480.4955F}
}

@ARTICLE{Ye+2020,
       author = {{Ye}, Claire S. and {Fong}, Wen-fai and {Kremer}, Kyle and {Rodriguez}, Carl L. and {Chatterjee}, Sourav and {Fragione}, Giacomo and {Rasio}, Frederic A.},
        title = "{On the Rate of Neutron Star Binary Mergers from Globular Clusters}",
      journal = {\apjl},
         year = 2020,
        month = jan,
       volume = {888},
       number = {1},
          eid = {L10},
        pages = {L10},
          doi = {10.3847/2041-8213/ab5dc5},
archivePrefix = {arXiv},
       eprint = {1910.10740},
 primaryClass = {astro-ph.HE},
       adsurl = {https://ui.adsabs.harvard.edu/abs/2020ApJ...888L..10Y}
}

@ARTICLE{Ivanova+2013,
       author = {{Ivanova}, N. and {Justham}, S. and {Chen}, X. and {De Marco}, O. and {Fryer}, C.~L. and {Gaburov}, E. and {Ge}, H. and {Glebbeek}, E. and {Han}, Z. and {Li}, X.-D. and {Lu}, G. and {Marsh}, T. and {Podsiadlowski}, P. and {Potter}, A. and {Soker}, N. and {Taam}, R. and {Tauris}, T.~M. and {van den Heuvel}, E.~P.~J. and {Webbink}, R.~F.},
        title = "{Common envelope evolution: where we stand and how we can move forward}",
      journal = {\aapr},
         year = 2013,
        month = feb,
       volume = {21},
          eid = {59},
        pages = {59},
          doi = {10.1007/s00159-013-0059-2},
archivePrefix = {arXiv},
       eprint = {1209.4302},
 primaryClass = {astro-ph.HE},
       adsurl = {https://ui.adsabs.harvard.edu/abs/2013A&ARv..21...59I}
}

@ARTICLE{Tauris+2017,
       author = {{Tauris}, T.~M. and {Kramer}, M. and {Freire}, P.~C.~C. and {Wex}, N. and {Janka}, H.-T. and {Langer}, N. and {Podsiadlowski}, Ph. and {Bozzo}, E. and {Chaty}, S. and {Kruckow}, M.~U. and {van den Heuvel}, E.~P.~J. and {Antoniadis}, J. and {Breton}, R.~P. and {Champion}, D.~J.},
        title = "{Formation of Double Neutron Star Systems}",
      journal = {\apj},
         year = 2017,
        month = sep,
       volume = {846},
       number = {2},
          eid = {170},
        pages = {170},
          doi = {10.3847/1538-4357/aa7e89},
archivePrefix = {arXiv},
       eprint = {1706.09438},
 primaryClass = {astro-ph.HE},
       adsurl = {https://ui.adsabs.harvard.edu/abs/2017ApJ...846..170T}
}

@ARTICLE{MarinPina+2025,
       author = {{Mar{\'\i}n Pina}, Daniel and {Gieles}, Mark and {Rastello}, Sara and {Iorio}, Giuliano},
        title = "{The formation of Gaia BH3}",
      journal = {arXiv e-prints},
         year = 2025,
        month = oct,
          eid = {arXiv:2510.07021},
        pages = {arXiv:2510.07021},
          doi = {10.48550/arXiv.2510.07021},
archivePrefix = {arXiv},
       eprint = {2510.07021},
 primaryClass = {astro-ph.GA},
       adsurl = {https://ui.adsabs.harvard.edu/abs/2025arXiv251007021M}
}

@ARTICLE{GaiaDR3,
       author = {{Gaia Collaboration} and {Vallenari}, A. and {Brown}, A.~G.~A. and {Prusti}, T. and {de Bruijne}, J.~H.~J. and {Arenou}, F. and {Babusiaux}, C. and {Biermann}, M. and {Creevey}, O.~L. and {Ducourant}, C. and {Evans}, D.~W. and {Eyer}, L. and {Guerra}, R. and {Hutton}, A. and {Jordi}, C. and {Klioner}, S.~A. and {Lammers}, U.~L. and {Lindegren}, L. and {Luri}, X. and {Mignard}, F. and {Panem}, C. and {Pourbaix}, D. and {Randich}, S. and {Sartoretti}, P. and {Soubiran}, C. and {Tanga}, P. and {Walton}, N.~A. and {Bailer-Jones}, C.~A.~L. and {Bastian}, U. and {Drimmel}, R. and {Jansen}, F. and {Katz}, D. and {Lattanzi}, M.~G. and {van Leeuwen}, F. and {Bakker}, J. and {Cacciari}, C. and {Casta{\~n}eda}, J. and {De Angeli}, F. and {Fabricius}, C. and {Fouesneau}, M. and {Fr{\'e}mat}, Y. and {Galluccio}, L. and {Guerrier}, A. and {Heiter}, U. and {Masana}, E. and {Messineo}, R. and {Mowlavi}, N. and {Nicolas}, C. and {Nienartowicz}, K. and {Pailler}, F. and {Panuzzo}, P. and {Riclet}, F. and {Roux}, W. and {Seabroke}, G.~M. and {Sordo}, R. and {Th{\'e}venin}, F. and {Gracia-Abril}, G. and {Portell}, J. and {Teyssier}, D. and {Altmann}, M. and {Andrae}, R. and {Audard}, M. and {Bellas-Velidis}, I. and {Benson}, K. and {Berthier}, J. and {Blomme}, R. and {Burgess}, P.~W. and {Busonero}, D. and {Busso}, G. and {C{\'a}novas}, H. and {Carry}, B. and {Cellino}, A. and {Cheek}, N. and {Clementini}, G. and {Damerdji}, Y. and {Davidson}, M. and {de Teodoro}, P. and {Nu{\~n}ez Campos}, M. and {Delchambre}, L. and {Dell'Oro}, A. and {Esquej}, P. and {Fern{\'a}ndez-Hern{\'a}ndez}, J. and {Fraile}, E. and {Garabato}, D. and {Garc{\'\i}a-Lario}, P. and {Gosset}, E. and {Haigron}, R. and {Halbwachs}, J.-L. and {Hambly}, N.~C. and {Harrison}, D.~L. and {Hern{\'a}ndez}, J. and {Hestroffer}, D. and {Hodgkin}, S.~T. and {Holl}, B. and {Jan{\ss}en}, K. and {Jevardat de Fombelle}, G. and {Jordan}, S. and {Krone-Martins}, A. and {Lanzafame}, A.~C. and {L{\"o}ffler}, W. and {Marchal}, O. and {Marrese}, P.~M. and {Moitinho}, A. and {Muinonen}, K. and {Osborne}, P. and {Pancino}, E. and {Pauwels}, T. and {Recio-Blanco}, A. and {Reyl{\'e}}, C. and {Riello}, M. and {Rimoldini}, L. and {Roegiers}, T. and {Rybizki}, J. and {Sarro}, L.~M. and {Siopis}, C. and {Smith}, M. and {Sozzetti}, A. and {Utrilla}, E. and {van Leeuwen}, M. and {Abbas}, U. and {{\'A}brah{\'a}m}, P. and {Abreu Aramburu}, A. and {Aerts}, C. and {Aguado}, J.~J. and {Ajaj}, M. and {Aldea-Montero}, F. and {Altavilla}, G. and {{\'A}lvarez}, M.~A. and {Alves}, J. and {Anders}, F. and {Anderson}, R.~I. and {Anglada Varela}, E. and {Antoja}, T. and {Baines}, D. and {Baker}, S.~G. and {Balaguer-N{\'u}{\~n}ez}, L. and {Balbinot}, E. and {Balog}, Z. and {Barache}, C. and {Barbato}, D. and {Barros}, M. and {Barstow}, M.~A. and {Bartolom{\'e}}, S. and {Bassilana}, J.-L. and {Bauchet}, N. and {Becciani}, U. and {Bellazzini}, M. and {Berihuete}, A. and {Bernet}, M. and {Bertone}, S. and {Bianchi}, L. and {Binnenfeld}, A. and {Blanco-Cuaresma}, S. and {Blazere}, A. and {Boch}, T. and {Bombrun}, A. and {Bossini}, D. and {Bouquillon}, S. and {Bragaglia}, A. and {Bramante}, L. and {Breedt}, E. and {Bressan}, A. and {Brouillet}, N. and {Brugaletta}, E. and {Bucciarelli}, B. and {Burlacu}, A. and {Butkevich}, A.~G. and {Buzzi}, R. and {Caffau}, E. and {Cancelliere}, R. and {Cantat-Gaudin}, T. and {Carballo}, R. and {Carlucci}, T. and {Carnerero}, M.~I. and {Carrasco}, J.~M. and {Casamiquela}, L. and {Castellani}, M. and {Castro-Ginard}, A. and {Chaoul}, L. and {Charlot}, P. and {Chemin}, L. and {Chiaramida}, V. and {Chiavassa}, A. and {Chornay}, N. and {Comoretto}, G. and {Contursi}, G. and {Cooper}, W.~J. and {Cornez}, T. and {Cowell}, S. and {Crifo}, F. and {Cropper}, M. and {Crosta}, M. and {Crowley}, C. and {Dafonte}, C. and {Dapergolas}, A. and {David}, M. and {David}, P. and {de Laverny}, P. and {De Luise}, F. and {De March}, R.},
        title = "{Gaia Data Release 3. Summary of the content and survey properties}",
      journal = {\aap},
         year = 2023,
        month = jun,
       volume = {674},
          eid = {A1},
        pages = {A1},
          doi = {10.1051/0004-6361/202243940},
archivePrefix = {arXiv},
       eprint = {2208.00211},
 primaryClass = {astro-ph.GA},
       adsurl = {https://ui.adsabs.harvard.edu/abs/2023A&A...674A...1G}
}

@article{Banerjee2026,
  title = {Stellar-mass black holes in young massive and open stellar clusters. VII. Comparisons with gravitational-wave events until LVK-O4a and Gaia compact binaries},
  author = {{Banerjee}, Sambaran},
  journal = {Phys. Rev. D},
  pages = {},
  year = {2026},
  month = {Sep},
  publisher = {American Physical Society},
  doi = {10.1103/q81s-7msp},
  url = {https://link.aps.org/doi/10.1103/q81s-7msp}
}

@ARTICLE{Kotko+2024,
       author = {{Kotko}, I. and {Banerjee}, S. and {Belczynski}, K.},
        title = "{The enigmatic origin of two dormant BH binaries: Gaia BH1 and Gaia BH2}",
      journal = {\mnras},
         year = 2024,
        month = dec,
       volume = {535},
       number = {4},
        pages = {3577-3594},
          doi = {10.1093/mnras/stae2591},
archivePrefix = {arXiv},
       eprint = {2403.13579},
 primaryClass = {astro-ph.SR},
       adsurl = {https://ui.adsabs.harvard.edu/abs/2024MNRAS.535.3577K}
}
\bibliographystyle{aasjournalv7.1}

\end{CJK*}
\end{document}